\documentclass[%
 aip,
rsi,%
 amsmath,amssymb,
reprint,%
]{revtex4-1}

\draft 

\usepackage[utf8]{inputenc} 
\usepackage[T1]{fontenc} 
\usepackage{physics} 
\usepackage{adjustbox} 
\usepackage{array} 
\usepackage[english]{babel} 
\usepackage{bbold, bbm, bm} 
\usepackage{booktabs} 
\usepackage{BOONDOX-uprscr}
\usepackage{braket} 
\usepackage{color} 
\usepackage{colortbl} 
\usepackage{csquotes} 
\usepackage{dcolumn} 
\usepackage{dsfont} 
\usepackage{enumitem} 
\usepackage{fancyhdr} 
\usepackage{float} 
\usepackage{gensymb} 
\usepackage{graphicx} 
\usepackage{hhline} 
\usepackage{makecell} 
\usepackage{mathalpha} 
\usepackage{mathrsfs} 
\usepackage{mathtools, amsthm} 
\usepackage{multirow,tabularx} 
\usepackage{ragged2e} 
\usepackage{setspace} 
\usepackage{soul} 
\usepackage{stmaryrd} 
\usepackage{tikz} 
\usepackage{xcolor} 
\usepackage{xspace} 
\usepackage{stackrel}
\usepackage{physics}
\usepackage{color,soul}

\newcolumntype{K}[1]{>{\centering\let\newline\\\arraybackslash\hspace{0pt}}m{#1}}
\newcolumntype{C}[1]{>{\centering\arraybackslash}p{#1}} 
\newcolumntype{Z}{>{\centering\arraybackslash}X} 

\newcommand{\wJ}{\omega_{J}\xspace} 
\newcommand{\wD}{\omega_{\Delta}\xspace} 
\newcommand{\thst}{\theta_\text{ST}} 
\newcommand{\Szero}{\ket{S_{0}}\xspace} 
\newcommand{\Tpos}{\ket{T_{+1}}\xspace} 
\newcommand{\Tzero}{\ket{T_{0}}\xspace} 
\newcommand{\Tneg}{\ket{T_{-1}}\xspace} 
\newcommand{\Szerostate}{\left({\ket{\alpha\beta}}-{\ket{\beta\alpha}}\right)/{\sqrt{2}}\xspace} 
\newcommand{\Tposstate}{\ket{\alpha\alpha}\xspace} 
\newcommand{\Tzerostate}{\left({\ket{\alpha\beta}}+{\ket{\beta\alpha}}\right)/{\sqrt{2}}\xspace} 
\newcommand{\Tnegstate}{\ket{\beta\beta}\xspace} 
\newcommand{\uL}[1]{#1~\mu \text{L}\xspace}

\newcommand{\thST}{\theta_\mathrm{ST}}
\newcommand{\Cth}{$\mathrm{^{13}C}$\xspace}
\newcommand{\Ctwo}{$\mathrm{^{13}C_2}$\xspace}
\newcommand{\F}[1]{$\mathrm{^{#1}F}$\xspace}

\newcommand{\wnut}{\omega_\mathrm{nut}}

\begin{document}


\title{Spinor Double-Quantum Excitation in the Solution NMR of Near-Equivalent Spin-1/2 Pairs}

\author{Urvashi D. Heramun}
\affiliation{School of Chemistry, University of Southampton, SO17 1BJ, UK}

\author{Mohamed Sabba}
\affiliation{School of Chemistry, University of Southampton, SO17 1BJ, UK}

\author{Dolnapa Yamano}
\affiliation{School of Chemistry, University of Southampton, SO17 1BJ, UK}

\author{Christian Bengs}
\affiliation{School of Chemistry, University of Southampton, SO17 1BJ, UK}

\author{Bonifac Legrady}
\affiliation{School of Chemistry, University of Southampton, SO17 1BJ, UK}

\author{Giuseppe Pileio}
\affiliation{School of Chemistry, University of Southampton, SO17 1BJ, UK}

\author{Sam Thompson}
\affiliation{School of Chemistry, University of Southampton, SO17 1BJ, UK}

\author{Malcolm H. Levitt}
\email[]{mhl@soton.ac.uk}
\affiliation{School of Chemistry, University of Southampton, SO17 1BJ, UK}

\date{\today}

\begin{abstract}
A family of double-quantum excitation schemes is described for the solution nuclear magnetic resonance (NMR) of near-equivalent spin-1/2 pairs. These new methods exploit the spinor behaviour of 2-level systems, whose signature is the change of sign of a quantum state upon a $2\pi$ rotation. The spinor behaviour is used to manipulate the phases of single-quantum coherences, in order to prepare a double-quantum precursor state which is rapidly converted into double-quantum coherence by a straightforward $\pi/2$ rotation. One set of spinor-based methods exploits symmetry-based pulse sequences, while the other set exploits SLIC (spin-lock-induced crossing), in which the nutation frequency under a resonant radiofrequency field is matched to the spin-spin coupling. A variant of SLIC is introduced which is well-compensated for deviations in the radiofrequency field amplitude. The methods are demonstrated by performing double-quantum-filtered \F{19} NMR on a molecular system containing a pair of diastereotopic \F{19} nuclei. The new methods are compared with existing techniques. 
\end{abstract}


\maketitle


\section{Introduction}
In high-field nuclear magnetic resonance (NMR), the term \emph{double-quantum coherence} (DQC)~\cite{EBW_1987} refers to a coherent superposition of spin states 
with quantum numbers for the angular momentum along the magnetic field differing by $\pm2$.
Signals passing through double-quantum coherence are easy to distinguish from other signals, due to their characteristic double-angle dependence on the radio-frequency phase. Double-quantum coherence is often used for signal selection and filtration~\cite{bax_natural_1980a,bax_nmr_1981}, for determining the connectivity of spin systems~\cite{bax_assignment_1981a,brouwer_symmetrybased_2005}, and for improving the resolution and phase properties of two-dimensional spectra~\cite{rance_DQFCOSY_1983}. Double-quantum relaxation is sensitive to cross-correlation effects, which convey motional and geometrical information~\cite{wokaun_MQ_1978}. 

Many different methods have been proposed for generating double-quantum coherence in both solid-state NMR~\cite{vega_fouriertransform_1976,lee_efficient_1995,brouwer_symmetrybased_2005} and solution NMR~\cite{bax_natural_1980a,bax_assignment_1981a,bax_nmr_1981,rance_DQFCOSY_1983}, starting from a spin ensemble in thermal equilibrium. The standard method for double-quantum excitation in solution NMR is a sequence of three strong radiofrequency pulses and two equal delays, as follows:
\begin{equation}
\label{eq:INADQ}
    \left(\frac{\pi}{2}\right)_y 
    - \tau_1 -
    \left(\pi\right)_x
    - \tau_1 -
    \left(\frac{\pi}{2}\right)_y. 
\end{equation}

This pulse sequence is generally known as INADEQUATE (Incredible Natural Abundance Double-Quantum Technique), since it was first used for the double-quantum filtration of \Cth NMR signals, allowing selective detection of the small signals from rare \Ctwo isotopologues~\cite{bax_natural_1980a,bax_assignment_1981a,bax_nmr_1981}. 

Consider an ensemble of homonuclear spin-1/2 pairs in isotropic solution. The system may be characterized by the parameters $\Delta\delta$ (the chemical shift difference between the members of the pair), and the isotropic coupling $J$ between the spins. The chemical shift frequency difference is given in angular units by $\wD=\omega^0\Delta\delta$, where the Larmor frequency in the magnetic field $B^0$ is given by $\omega^0=-\gamma B^0$, and $\gamma$ is the magnetogyric ratio. The chemical shift frequency difference is given in Hz by $\Delta=\wD/(2\pi)$. 

\begin{figure}[bt]
\centering
\includegraphics[width=0.5\textwidth]{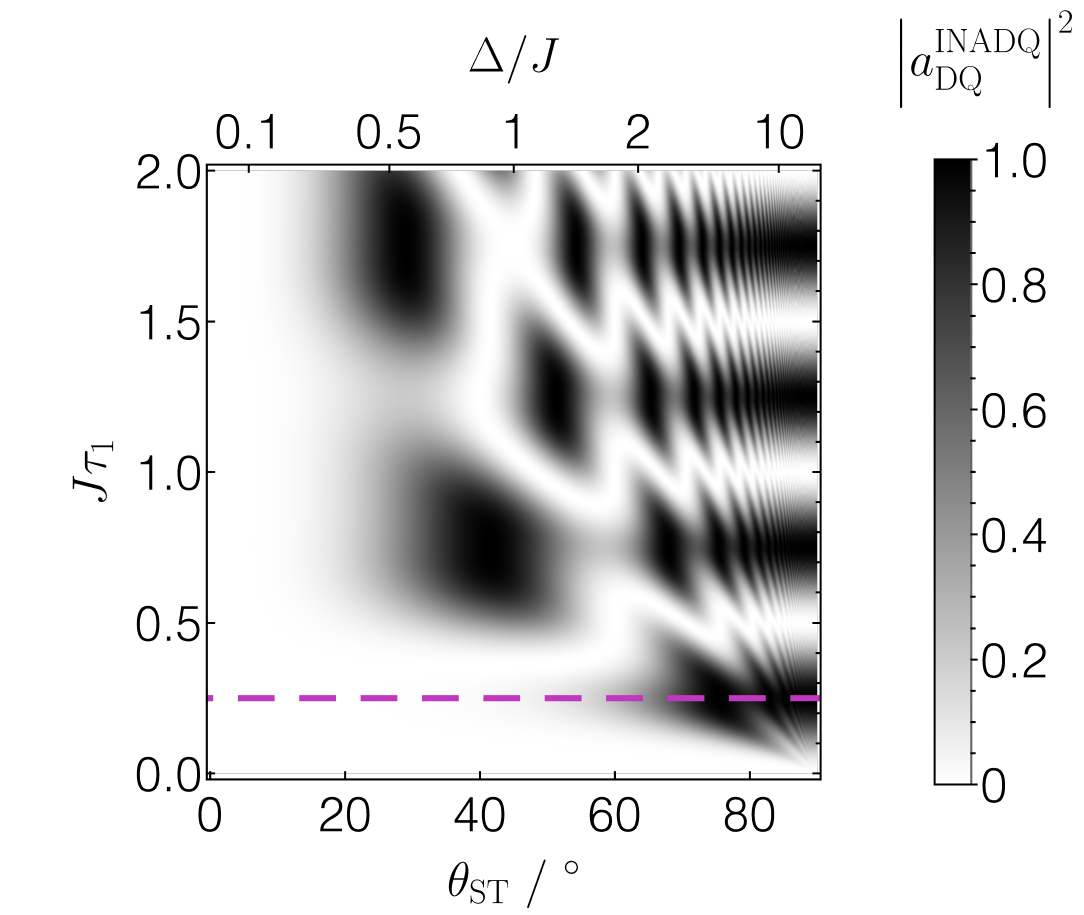}
\caption{Double-quantum filtering amplitude for the INADEQUATE sequence of Equation~\ref{eq:INADQ}. The square magnitude of the double-quantum excitation amplitude, $\vert  a_\mathrm{DQ}^\mathrm{INADQ}\vert^2$, is plotted against the singlet-triplet mixing angle $\thST$ (horizontal) and the interpulse delays $\tau_1$ (vertical), using the expression in Equation~\ref{eq:aDQINADQ}. The top axis shows the values of the ratio $\Delta/J$. Near-ideal double-quantum excitation is achieved for $\tau_1=(4J)^{-1}$ (horizontal dashed line), but only in the weakly-coupled regime  $\thST\gtrsim70\degree$. Note the weak double-quantum excitation in the near-equivalence regime ($\thST\lesssim 30 \degree$).}
\label{fig:INADQ-density-plot}
\end{figure} 

The coupling regime of the spin-1/2 pair is conveniently defined by the 
\emph{singlet-triplet mixing angle} $\thST$, defined by~\cite{bengs_gM2S_2020}:
\begin{equation}
\label{eq:thST}
\thST = \arctan(\wD/\wJ) = \arctan(\Delta/J).
\end{equation}
Systems with $\thST\gtrsim70\degree$ are said to be \emph{weakly coupled}. Systems with $\thST\lesssim70\degree$ are said to be \emph{strongly coupled}. The extreme strong-coupling regime $\thST\lesssim30\degree$ is the main focus of the present article. This is called \emph{near-equivalence}~\cite{tayler_singlet_2011,stevanato_longlived_2015}.  

The three-pulse INADEQUATE sequence of Equation~\ref{eq:INADQ} operates well in the weak-coupling regime $\thST\gtrsim70\degree$, where setting the delays to $\tau_1=(4J)^{-1}$ leads to good double-quantum excitation, independent of the shift difference $\Delta$. However, the INADEQUATE sequence lives up to its self-deprecating moniker outside the weak-coupling regime. Several authors~\cite{bax_investigation_1980,kay_product_1988,nakai_inadequate_1993,wilman_doublequantum_1994} have independently derived an expression for the double-quantum excitation amplitude of INADEQUATE in the general case. This may be written as follows:
\begin{multline}
\label{eq:aDQINADQ}
a_\mathrm{DQ}^\mathrm{INADQ} =
i \cos(\thST)  
\sin(\Omega_\mathrm{ST} \tau_1)
\cos(\wJ\tau_1)
\\
 -i \left(
\sin^2(\thST)+\cos^2(\thST)
\cos(\Omega_\mathrm{ST} \tau_1)
\right)\sin(\wJ\tau_1),
\end{multline}
where:
\begin{align}
\label{eq:symbols}
\wJ &= 2\pi J,
\nonumber\\
\wD &= 2\pi\Delta,
\nonumber\\
\Omega_\mathrm{ST} &= (\wJ^2 + \wD^2)^{1/2}.
\end{align}

Figure~\ref{fig:INADQ-density-plot} shows a density plot of the squared double-quantum excitation amplitude $\vert  a_\mathrm{DQ}^\mathrm{INADQ}\vert^2$ as a function of the angle $\thST$ and the interpulse delay $\tau_1$. Ideal double-quantum excitation is achieved in the weak-coupling case $\thST\gtrsim70\degree$ by setting the interpulse intervals to $\tau_1=|4J|^{-1}$ 
(horizontal dashed line in Figure~\ref{fig:INADQ-density-plot}).
However, double-quantum excitation is poor in the strong-coupling case ($\thST\lesssim70\degree$), unless the interpulse delays $\tau_1$ are greatly extended and carefully chosen~\cite{bax_investigation_1980,nakai_inadequate_1993}. In practice, relaxation losses usually lead to poor efficiency, especially in the case of near-equivalence ($\thST\lesssim30\degree$). 

\emph{Geometric double-quantum excitation (GeoDQ)} provides efficient double-quantum excitation in the near-equivalence regime, by exploiting the geometric Aharanov-Anandan phase~\cite{aharonov_phase_1987} in a zero-quantum subspace~\cite{bengs_aharonov_2023}. Geometric double-quantum excitation has been used to selectively detect the signals from near-equivalent \Ctwo pairs in \Cth NMR spectra~\cite{heramun_singlet_2025}.

However, despite its efficiency, the set-up of the GeoDQ sequence requires detailed knowledge of the spin-system parameters, and it is not easy to optimise for an unknown spin system.

In this article, we introduce a set of methods which are based on a different principle to the GeoDQ technique, but which also achieve efficient double-quantum excitation in the near-equivalence regime. We call these methods \emph{Spinor-DQ excitation} since they exploit the spinor property of two-level quantum systems -- namely that the rotation of a quantum state by $2\pi$ restores the original state, but with a sign change~\cite{cartan_theory_1981,aharonov_observability_1967}. Several magnetic resonance experiments have demonstrated spinor behaviour~\cite{stoll_explicit_1977,wolff_spinor_1979,mehring_blochsiegert_1986,suter_indirect_1986,thrippleton_startmas_2006,thrippleton_satellite_2008}. In the current work, we exploit spinor behaviour to generate double-quantum coherence in ensembles of spin-1/2 pairs, in the near-equivalence regime. 

\begin{figure}[b]
\centering
\includegraphics[width=0.75\columnwidth]{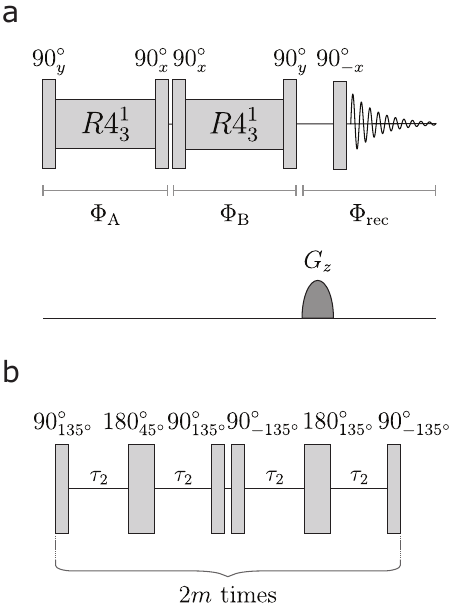}
\caption{
(a) Double-quantum-filtering pulse sequence, using PulsePol/$R4_3^1$ for double-quantum excitation, including a $z$-filtering step before signal acquisition. The phases $\Phi_\mathrm{A}$, $\Phi_\mathrm{B}$, and $\Phi_\mathrm{rec}$ are cycled in 4 steps to implement double-quantum filtering.
(b) Structure of the $R4_3^1$ sequence, as shown in Equation~\ref{eq:R431}.
}
\label{fig: pulsepol dq pulse sequence}
\end{figure} 

We demonstrate two different implementations of Spinor-DQ excitation: One implementation, shown in Figure~\ref{fig: pulsepol dq pulse sequence}(a), uses the \emph{PulsePol} sequence, which was originally developed for electron-nucleus polarization transfer in diamond nitrogen-vacancy systems~\cite{schwartz_robust_2018}. It has been shown that the PulsePol sequence may also be used for the solution NMR spectroscopy of spin-1/2 pairs~\cite{tratzmiller_pulsed_2021,sabba_symmetrybased_2022,harbor-collins_103rh_2023,harbor-collins_nmr_2024,harbor-collins_1henhanced_2024}, and that its operation may be understood in terms of symmetry-based recoupling theory, as originally developed for magic-angle-spinning solid-state NMR~\cite{lee_efficient_1995,carravetta_symmetry_2000,levitt_symmetrybased_2007,levitt_symmetry_2008}. 
The notation $R4_3^1$, used in Figure~\ref{fig: pulsepol dq pulse sequence}(a), emphasises this connection, which is explained further below. 

\begin{figure}[tb]
\centering
\includegraphics[width=0.75\columnwidth]{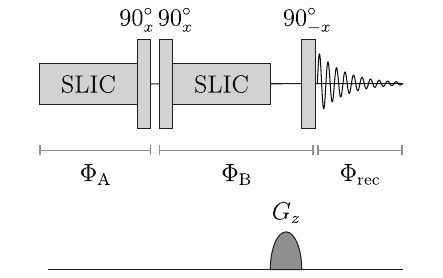}
\caption{
Double-quantum-filtering pulse sequence, using SLIC for double-quantum excitation. A $z$-filtering step is included before acquisition. The phases $\Phi_\mathrm{A}$, $\Phi_\mathrm{B}$ and $\Phi_\mathrm{rec}$ are cycled in 4 steps to implement double-quantum filtering.
The SLIC element is applied with phase $\phi=0$, such that the SLIC field has the \emph{same phase} as the subsequent $90\degree$ pulse. The compensated cSLIC sequence may be used as the SLIC element, as described in Section~\ref{sec:cSLIC}.
}
\label{fig: slic dq Gz pulse sequence}
\end{figure} 

A second implementation of Spinor-DQ excitation uses the 
spin-lock-induced crossing (SLIC) method, in which a resonant radiofrequency field is applied such that the nutation frequency of the nuclear spins matches the $J$-coupling between the members of the spin pair: $\omega_\mathrm{nut}=\omega_J$~\cite{devience_preparation_2013a,
devience_nuclear_2013,
elliott_longlived_2016,
elliott_longlived_2016a,
eills_singlet_2017,
sheberstov_cis_2018,
elliott_nuclear_2019,
sheberstov_excitation_2019,
sheberstov_generating_2019,
devience_manipulating_2020,
korenchan_31p_2022,
sonnefeld_polychromatic_2022,
sonnefeld_longlived_2022,
razanahoera_hyperpolarization_2024,
sheberstov_collective_2024,
wiame_longlived_2025,
mandzhieva_zerofield_2025,
mcbride_scalable_2025}.

The general scheme for the SLIC implementation of double-quantum filtering is shown in figure~\mbox{\ref{fig: slic dq Gz pulse sequence}}. In this case, note the unusual \emph{absence} of an initial $90\degree$ pulse before the SLIC pulse is applied to a state of longitudinal magnetization. Furthermore, note that the $90\degree$ pulse which follows the SLIC pulse has the \emph{same phase} as the SLIC field.
This contrasts to the use of SLIC to generate singlet order, and for similar applications, where an initial $90\degree$ pulse, with a $90\degree$ phase shift relative to the SLIC field, is included\cite{devience_preparation_2013a,
devience_nuclear_2013,
elliott_longlived_2016,
elliott_longlived_2016a,
eills_singlet_2017,
sheberstov_cis_2018,
elliott_nuclear_2019,
sheberstov_excitation_2019,
sheberstov_generating_2019,
devience_manipulating_2020,
korenchan_31p_2022,
sonnefeld_polychromatic_2022,
sonnefeld_longlived_2022,
razanahoera_hyperpolarization_2024,
sheberstov_collective_2024,
wiame_longlived_2025,
mandzhieva_zerofield_2025,
mcbride_scalable_2025}.

As described below, the use of SLIC to generate double-quantum coherence, using the pulse sequence shown in Figure~\ref{fig: slic dq Gz pulse sequence}, is particularly sensitive to small deviations in the rf field amplitude. In Section~\ref{sec:cSLIC}, we introduce a compensated variant called cSLIC, which employs radiofrequency pulses with two different amplitudes. cSLIC is much more robust than standard SLIC, with respect to deviations in the radiofrequency field amplitude. 

The rest of this paper is organized as follows: The general quantum theory of near-equivalent spin-1/2 pair systems is given in section~\ref{sec:Theory}.
Section~\ref{sec:GeoDQ} reviews the principles of geometric DQ excitation, since an appreciation of how this method works is informative for the Spinor-DQ techniques.
Section~\ref{sec:Spinor-DQ} describes the principles of Spinor-DQ excitation, including both the PulsePol/Symmetry-Based and the SLIC implementations. The rf-compensated cSLIC method is also described here. 
Sections~\ref{sec:Methods} and \ref{sec:Results} present results for a demonstration of double-quantum excitation using
the \F{19} solution NMR of a molecular system containing a diastereotopic pair of \F{19} nuclei. 
Both Spinor-DQ and GeoDQ procedures are shown to excite double-quantum coherence in the ensemble of near-equivalent \F{19} spin pairs, with much higher efficiency than the conventional INADEQUATE sequence.
The paper concludes by discussing some possible applications of double-quantum excitation in the solution NMR of near-equivalent spin systems. 

\section{
Theoretical Background}
\label{sec:Theory}

\subsection{Spin Hamiltonian}
\label{sec:Hamiltonian}
The rotating-frame spin Hamiltonian for a homonuclear spin-1/2 pair in the absence of a radiofrequency field is as follows:
\begin{equation}
H^0 = \Omega_1 I_{1z} + \Omega_2 I_{2z} + H_J,
\end{equation}
where the $J$-coupling Hamiltonian is given by:
\begin{equation}
H_{J} = \wJ\,\mathbf{I}_{1}\cdot\mathbf{I}_{2}, 
\end{equation}
and the resonance offset frequencies for the two spins are defined as follows:
\begin{align}
\Omega_1 
&= 
\omega^0 (\delta_1 - \delta_\mathrm{ref}),
\nonumber\\
\Omega_2 &= 
\omega^0 (\delta_2 - \delta_\mathrm{ref}).
\end{align}
The terms $\{\delta_1, \delta_2, \delta_\mathrm{ref}\}$ are the chemical shift of spins $I_1$, spin $I_2$, and the rf reference frequency, respectively. The other symbols are specified in Equation~\ref{eq:symbols}.

The spin Hamiltonian may be written as the sum of two commuting terms, as follows:
\begin{equation}
\label{eq:H0}
H^0 = H_{\Sigma} + 
(H_{\Delta} + H_{J}).
\end{equation}
where:
\begin{align}
\label{eq:H0terms}
H_{\Sigma} &= 
\omega_{\Sigma} 
\,\tfrac{1}{2}
\left(I_{1z} + I_{2z}\right), 
\nonumber\\
H_{\Delta} &= 
\omega_{\Delta} 
\,\tfrac{1}{2}
\left(I_{1z} - I_{2z}\right).
\end{align}
The sum and difference of the resonance offset frequencies are defined as follows:
\begin{align}
\omega_{\Sigma} &= \Omega_1 + \Omega_2,
\nonumber\\
\omega_{\Delta} &= \Omega_1 - \Omega_2.
\end{align}

The singlet and triplet states of the spin-1/2 pair are defined and numbered as follows:
\begin{alignat}{3}
\label{eq:States}
\ket{1} &= \Szero &=& \Szerostate,
\nonumber\\
\ket{2} &=\Tpos &=& \Tposstate,
\nonumber\\
\ket{3} &=\Tzero &=& \Tzerostate,
\nonumber\\
\ket{4} &=\Tneg &=& \Tnegstate.
\end{alignat}
Here $\alpha$ and $\beta$ denote the two spin angular momentum projections $\pm (\hbar/2)$  along an external axis, and the subscripts $M_I\in\{+1,0,-1\}$ refer to the value of the magnetic spin quantum number. The numbering system of Equation~\ref{eq:States} is used throughout the rest of this article.

Single-transition operators for an arbitrary two-level subspace of the states $\ket{r}$  and $\ket{s}$ are defined as follows~\cite{vega_fictitious_1978,wokaun_selective_1977}:
\begin{align}
I_{x}^{rs} &= \tfrac{1}{2} \left(\ket{r}\bra{s} + \ket{s}\bra{r}\right), 
\nonumber \\[5pt]
I_{y}^{rs} &= \tfrac{1}{2i} \left(\ket{r}\bra{s} - \ket{s}\bra{r}\right), 
\nonumber \\[5pt]
I_{z}^{rs} &= \tfrac{1}{2} \left(\ket{r}\bra{r} - \ket{s}\bra{s}\right), 
\nonumber \\[5pt]
\mathbb{1}^{rs} &= \ket{r}\bra{r} + \ket{s}\bra{s} .
\end{align}

Rotation operators for single transitions are defined as follows:
\begin{equation}
\label{eq:Rrs}
R_{\mu}^{rs}\left(\beta\right) 
= \exp\!\left(-i \beta I_{\mu }^{rs} \right),
\end{equation}
where $\mu\in\{x,y,z\}$ denotes the axis of rotation. The operator for a cycle, i.e. a rotation through an angle of $2\pi$, is particularly relevant. A cycle in the $\left\{\ket{r},\ket{s} \right\}$ subspace is independent of the axis, and is described by the operator:
\begin{equation}
\mathcal{C}^{rs} = R_{\mu}^{rs}(2\pi),
\end{equation}
which may be written as follows:
\begin{equation}
  \mathcal{C}^{rs} =
  -\mathbb{1}^{rs}+\sum_{u\neq\{r,s\}}\ket{u}\bra{u},
\end{equation}
where the second term involves a summation over  all states which are outside the rotated 2-level system. For example, the cycle operator for the $\{\ket{1},\ket{2}\}$ subspace of the 4-level system is given by:
\begin{equation}
\label{eq:C12}
    C^{12} = -\mathbb{1}^{12}+\mathbb{1}^{34}.
\end{equation}
The negative sign of the first term is a consequence of spinor behaviour. 

The theory of Spinor-DQ excitation is developed most conveniently by expressing the 
spin Hamiltonian $H^0$ 
in terms of single-transition operators, as follows:
\begin{align}
H^0 &=
-\Omega_\mathrm{ST} R_{y}^{13}(-\thST)I_z^{13}
            R_{y}^{13}(+\thST)
+\omega_\Sigma I_z^{24}
\nonumber\\
   & \qquad +\tfrac{1}{4}\wJ (\mathbb{1}^{24}-\mathbb{1}^{13}),
\end{align}
where $\Omega_\mathrm{ST}$ and $\thST$ are given in Equations~\ref{eq:thST} and \ref{eq:symbols}. 

\begin{figure}[tb] \centering 
\includegraphics[width=0.85\columnwidth]{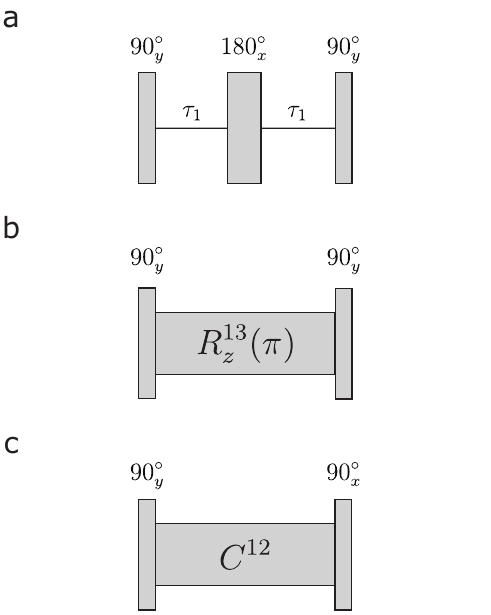} 
\caption{
Double-quantum excitation schemes for spin-1/2 pairs. (a) The INADEQUATE three-pulse method.
(b) The geometric double-quantum method. The element $R_z^{13}(\pi)$ indicates a $\pi$ rotation about the $z$-axis of the zero-quantum subspace, as described in Refs.\cite{bengs_aharonov_2023,heramun_singlet_2025}.
(c) The PulsePol/symmetry-based} implementation of the spinor double-quantum-excitation method. The element $C^{12}$ indicates a $2\pi$ rotation in the $\{\ket{1},\ket{2}\}$ singlet-triplet subspace, generated by a symmetry-based pulse sequence. 
\label{fig:DQexc-schemes}
\end{figure} 

\subsection{Double-Quantum Excitation and Double-Quantum Filtering}
\label{sec:DQexcitation}

\subsubsection{Double-quantum excitation amplitude}
All double-quantum excitation schemes considered here act upon a spin-pair ensemble in thermal equilibrium, in the presence of a strong magnetic field $B^0$ along the $z$-axis of the laboratory coordinate frame. This equilibrium state is described by a spin density operator with excess population in the $\ket{2}=\ket{T_{+1}}$ state, and depleted population in the $\ket{4}=\ket{T_{-1}}$ state (assuming positive gyromagnetic ratio $\gamma$). The thermal equilibrium density operator may be written as follows (omitting the unimportant $\mathbb{1}$ operator component, and numerical factors):
\begin{equation}
\label{eq:rho1}
\rho_1 = I_z = 2I_z^{24}.
\end{equation}

If dissipative losses are ignored, double-quantum excitation generates a unitary transformation $U$ of the density operator, where $U^\dagger=U^{-1}$. The \emph{double-quantum excitation amplitude} $a_\mathrm{DQ}$ for a particular scheme is defined as follows:
\begin{equation}
\label{eq:aDQ}
 a_\mathrm{DQ} = 
    \bra{2}{U \rho_1 U^\dagger}\ket{4}.
\end{equation}
This may also be written as follows:
\begin{equation}
\label{eq:aDQ}
 a_\mathrm{DQ} = 
    \big\langle I_z 
    \overset{U}\to
    \ket{2}\bra{4}\big\rangle,
\end{equation}
where the \emph{transformation amplitude}~\cite{eills_singlet_2017} is defined as following:
\begin{equation}
\label{eq:TransformationAmplitude}
\langle A \overset{U}\to B\rangle
=\frac{(B|UAU^\dagger)}{(B|B)}
\end{equation}
The Liouville bracket~\cite{jeener_superoperators_1982} is given by:
\begin{equation}
    (B|A) = \mathrm{Tr}\{B^\dagger A\}.
\end{equation}
The \emph{transformation amplitude} $\langle A \overset{U}\to B\rangle$ represents the amplitude for conversion of operator $A$ into operator $B$ by the unitary transformation $U$. The \emph{SpinDynamica}~\cite{bengs_spindynamica_2018} routine \texttt{TransformationAmplitude} implements this definition.

\subsubsection{Double-quantum filtration amplitude}
Double-quantum coherence is not associated with net magnetization and does not generate an observable NMR signal. In most double-quantum experiments, the excited double-quantum coherence is reconverted into observable magnetization by applying a repeat of the excitation sequence, in some cases in reverse chronological order. Signals which do not pass through double-quantum coherence at the junction of the excitation and reconversion sequences are suppressed by a standard 4-step phase cycling procedure, as described in Section~\ref{sec:Instrumental}. This method is called \emph{double-quantum filtration (DQF)}. In most cases, the overall signal amplitude, after double-quantum filtration, is equal to the square magnitude of the double-quantum excitation amplitude (apart from a possible sign change)~\cite{weitekamp_timedomain_1982,eills_singlet_2017}:
\begin{multline}
\label{eq:aDQF}
a_\mathrm{DQF}=
    |a_\mathrm{DQ}|^2
=|\bra{2}{U \rho_1 U^\dagger}\ket{4}|^2
=|\big\langle \rho_1 \overset{U}{\to} \ket{2}\bra{4}\big\rangle|^2.
\\
\end{multline}
Although $a_\mathrm{DQ}$ is a complex number in general, the DQ filtering amplitude $a_\mathrm{DQF}$ is real and positive. The amplitude $|a_\mathrm{DQF}|$ approaches $1$ for a scheme of maximum efficiency. 

For all the schemes discussed below, the symbol $T$ is used for the total duration of the double-quantum excitation sequence. In most cases, $T$ should be as small as possible, in order to minimize relaxation losses. In practice, the performance of a sequence is a compromise between higher theoretical efficiency in the absence of relaxation, and relaxation losses associated with a large value of $T$. This is particularly the case for near-equivalent systems, where large values of $a_\mathrm{DQF}$ require large values of $T$. 

\begin{figure}[tb]
\centering
\includegraphics[width=0.75\columnwidth]{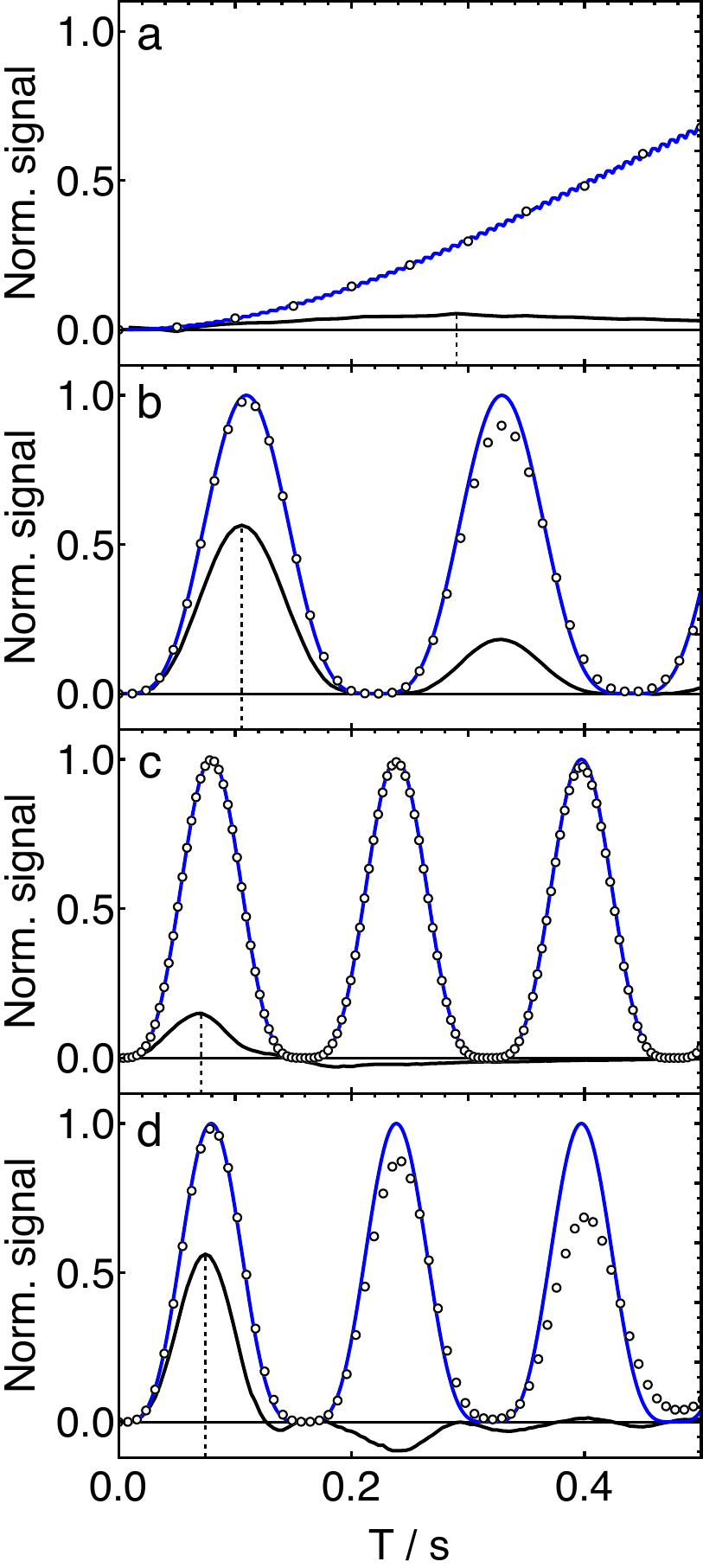}
\caption{
Dependence of double-quantum-filtered signals on sequence duration $T$ for different excitation schemes:
Analytical functions neglecting relaxation (blue curves), simulated points neglecting relaxation (open circles), and experimental data for the solution of \textbf{I} (solid black lines).
(a): INADEQUATE (blue curve: $|a_\mathrm{DQ}^\mathrm{INADQ}(T)|^2$, where $a_\mathrm{DQ}^\mathrm{INADQ}(T)$ is given by equation~\ref{eq:aDQINADQ});
(b): PulsePol-DQ sequence (blue curve: equation~\ref{eq:aDQFpulsepol});
(c): SLIC-DQ sequence, using the implementation in Figure~\ref{fig:SLIC-DQ-sequences}(b) (blue curve: equation~\ref{eq:aDQFSLICb});
(d): cSLIC-DQ sequence (blue curve: equation~\ref{eq:aDQFcSLIC}). All simulations use $J = 255.94$~Hz and $\Delta=17.8$~Hz. The experimentally optimised values of $T$ are shown by vertical dashed lines. 
}
\label{fig:DQtrajectories}
\end{figure}

\subsubsection{INADEQUATE}
The INADEQUATE method is given by Equation~\ref{eq:INADQ}, as shown in Figure~\ref{fig:DQexc-schemes}(a). 
The total duration is $T=2\tau_1$ in the limit of infinitely short radiofrequency pulses. 
INADEQUATE leads to the double-excitation amplitude of Equation~\ref{eq:aDQINADQ}, which is plotted in Figure~\ref{fig:INADQ-density-plot}. 
The theoretical dependence of $a_\mathrm{DQF}$ on excitation duration $T$ for INADEQUATE is shown by the blue line in Figure\ref{fig:DQtrajectories}(a), using the spin-system and experimental parameters defined in section\ref{sec:Methods}. This illustrates the very slow build-up of double-quantum coherence for INADEQUATE in the near-equivalence regime.

Although the standard INADEQUATE sequence performs well for weakly-coupled spin-1/2 pairs, significant double-quantum excitation is only achieved in the near-equivalence regime when the total duration $T$ of the INADEQUATE is of the order of the inverse of the inner peak splitting:
\begin{equation}
    T \sim   \vert\pi/\Delta\omega_\mathrm{inner}\vert
    \simeq 
    \vert  J/\Delta^2 \vert.
\end{equation}
This leads to a large value of $T$ in the near-equivalence limit, and correspondingly large relaxation losses.

Much faster double-quantum excitation is achieved in the near-equivalence regime by two different approaches: geometric double-quantum excitation 
and spinor double-quantum excitation.

\begin{figure}[tb]
\centering
\includegraphics[width=0.85\columnwidth]{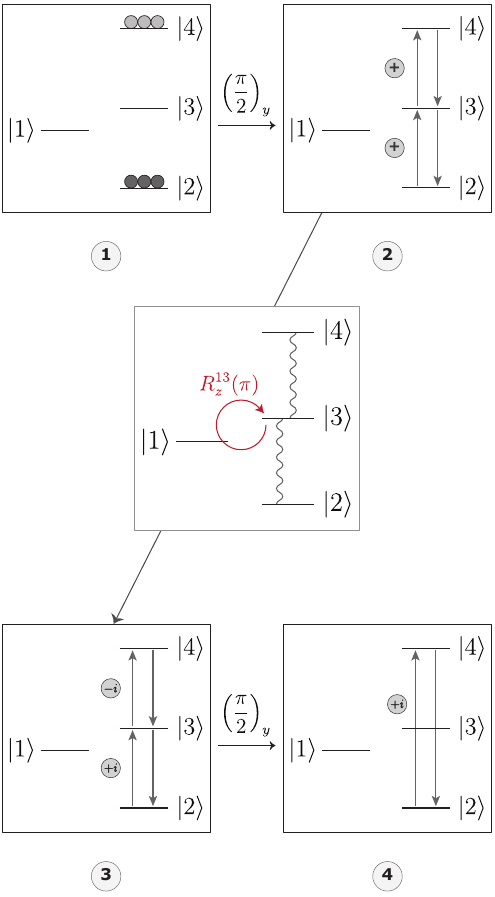}
\caption{
Geometric double-quantum excitation.
An ensemble of spin-1/2 pairs is at thermal equilibrium in a strong magnetic field (pane 1). In-phase single-quantum triplet-triplet coherences are generated by a $\pi/2$ $y$-pulse (pane 2). This is converted into a \emph{double-quantum precursor state} of antiphase triplet-triplet coherences (pane 3). A final $\pi/2$ $y$-pulse generates double-quantum coherence between the states $\ket{T_{\pm1}}$ (pane 4). The transformation from the state in pane 2 to that in pane 3 is achieved by generating a cyclic trajectory with geometric phase $\pi$ in the zero-quantum subspace spanned by the singlet state $\Szero=\ket{1}$ and the central triplet state $\Tzero=\ket{3}$. This transformation corresponds to a rotation through $\pi$ about the $z$-axis of the zero-quantum subspace. 
}
\label{fig:DQ-excitation-GeoDQ}
\end{figure} 

\section{Geometric double-quantum excitation}
\label{sec:GeoDQ}

\subsection{General principles}

The principles of geometric DQ excitation\cite{bengs_aharonov_2023} are reviewed here, using a different theoretical approach to that taken in Ref.\cite{bengs_aharonov_2023}. This allows a direct correspondence to be drawn between GeoDQ and Spinor-DQ excitation. 

A general scheme for geometric DQ excitation is shown in figure~\ref{fig:DQexc-schemes}(b). Figure~\ref{fig:DQ-excitation-GeoDQ} depicts the corresponding density operator transformations in graphical form.

\subsubsection{Initial state}
The initial state $\rho_1$ (equation~\ref{eq:rho1}) is depicted in pane 1, using darker balls to represent enhanced population, and lighter balls to represent depleted populations. 

\subsubsection{Single-quantum excitation}
The GeoDQ sequence starts with the application of a strong, resonant, $\pi/2$ pulse along the rotating-frame $y$-axis, as shown in figure~\ref{fig:DQexc-schemes}(b). This pulse generates single-quantum coherences between adjacent pairs of triplet states:
\begin{equation}
\rho_2 =
 R_y(\pi/2)\rho_1 R_y(\pi/2)^\dagger = I_x.
 \end{equation}
This operator may be written in terms of the triplet single-transition operators as follows:
\begin{align}
\label{eq:rho2}
\rho_2 &= I_x  = 2^{1/2}(I_{x}^{23}+I_{x}^{34})
\nonumber\\
&= 2^{-1/2}\big\{
\ket{2}\bra{3}+\ket{3}\bra{2}
+\ket{3}\bra{4}+\ket{4}\bra{3}
\big\}.
\nonumber\\
\end{align}
The single-quantum triplet-triplet coherences $\ket{T_0}\bra{T_{-1}}=\ket{3}\bra{4}$, $\ket{T_{+1}}\bra{T_{0}}=\ket{2}\bra{3}$, 
$\ket{T_{-1}}\bra{T_0}=\ket{4}\bra{3}$ and
$\ket{T_{0}}\bra{T_{+1}}=\ket{3}\bra{2}$ are all generated with the same sign. This is depicted by the four arrows in pane 2 of Figure~\ref{fig:DQ-excitation-GeoDQ}, and the adjacent "+" signs. 

\subsubsection{Preparation of the double-quantum precursor}
In order to excite double-quantum coherence, the state $\rho_2$, which has the same signs for all single-quantum triplet-triplet coherences, must be converted into a state $\rho_3$ with opposite signs for pairs of single-quantum triplet-triplet coherence. This antiphase single-quantum state is called here the \emph{double-quantum precursor state}.

In the near-equivalence limit, the key transformation is accomplished as follows:
A sequence of $\pi$ pulses with carefully selected timings generates a cyclic trajectory in the zero-quantum (ZQ) subspace spanned by the singlet state $\ket{1}=\ket{S_0}$ and the central triplet state $\ket{3}=\ket{T_0}$, as depicted in the central pane of Figure~\ref{fig:DQ-excitation-GeoDQ}. As discussed in Ref.~\cite{bengs_aharonov_2023}, the cyclic trajectory encloses a solid angle of $\pi$ about the origin of the zero-quantum subspace, which leads to a geometric Aharanov-Anandan phase of $\pi/2$ for the states $\ket{1}$ and $\ket{3}$. The net effect is equivalent to a $R_z^{13}(\pi)$ rotation, as shown in figure~\ref{fig:DQexc-schemes}(b). 

The triplet-triplet coherences $\ket{3}\bra{4}$ and $\ket{2}\bra{3}$ acquire opposite phase shifts of $\pm\pi/2$ upon completion of the cyclic zero-quantum trajectory, leading to a double-quantum precursor state with triplet-triplet coherences of opposite sign (see pane 3 of Figure~\ref{fig:DQ-excitation-GeoDQ}):
\begin{align}
\label{eq:rho3geoDQ}
\rho_3^\mathrm{geo} &=
 R_z^{13}(\pi)\,\rho_2\, R_z^{13}(\pi)^\dagger
 \nonumber\\
 &= 2^{1/2}R_z^{13}(\pi)
 \left(
 I_{x}^{23}+I_{x}^{34}
 \right)
 R_z^{13}(\pi)^\dagger
 \nonumber\\
 &= 2^{1/2}
 \left(
 I_{y}^{23}-I_{y}^{34}
 \right)
 \nonumber\\
&= 2^{-1/2}\,i\big\{
-\ket{2}\bra{3}+\ket{3}\bra{2}
 \nonumber\\&
\qquad\qquad\qquad
+\ket{3}\bra{4}-\ket{4}\bra{3}
\big\}.
\end{align}
This transformation follows from the commutation relationships of single-transition operators~\cite{vega_fictitious_1978,wokaun_selective_1977}:
\begin{align}
[I_z^{13},I_x^{23}] &= \tfrac12 i I_y^{23},
\nonumber\\
[I_z^{13},I_y^{23}] &= - \tfrac12 i I_x^{23},
\nonumber\\
[I_z^{13},I_x^{34}] &= - \tfrac12 i I_y^{34},
\nonumber\\
[I_z^{13},I_y^{34}] &= \tfrac12 i I_x^{34}.
\end{align}
The density operator of Equation~\ref{eq:rho3geoDQ} may be written in terms of Cartesian product operators~\cite{nakai_inadequate_1993} 
as follows:
\begin{equation}
\label{eq:rho3geoCart}
\rho_3^\mathrm{geo} = 2I_{1y}I_{2z}+2I_{1z}I_{2y}.
\end{equation}

\subsubsection{Double-quantum excitation}
The double-quantum precursor state $\rho_3^\mathrm{geo}$ is converted into double-quantum coherence by a second $\pi/2$ pulse with phase $\pi/2$, as shown in figure~\ref{fig:DQexc-schemes}(b). 

The relevant transformation is as follows:
\begin{align}
\rho_4 
&=R_y(\pi/2)\,\rho_3^\mathrm{geo}\, R_y(\pi/2)^\dagger
 \nonumber\\
 &=R_y(\pi/2)\,
 (2I_{1y}I_{2z}+2I_{1z}I_{2y})
  R_y(\pi/2)^\dagger
\nonumber\\
&=
 2I_{1x}I_{2y}+2I_{1y}I_{2x}.
\end{align}

The resulting double-quantum state may be expressed as follows:
\begin{align}
\label{eq:rho4}
\rho_4 
&=
 2I_{1x}I_{2y}+2I_{1y}I_{2x}
\nonumber\\
 &=
 -i(I_1^+ I_2^+ - I_1^- I_2^-)
 \nonumber\\
 &=
 -i(\ket{2}\bra{4}-\ket{4}\bra{2} ) = 2I_y^{24}.
\end{align}
A graphical representation is shown in pane 4 of Figure~\ref{fig:DQ-excitation-GeoDQ}.

The final state corresponds to double-quantum excitation of maximal amplitude:
\begin{align}
    a_\mathrm{DQ} &= \bra{2}\rho_4\ket{4} = -i,
\nonumber\\
    a_\mathrm{DQ}^* &= \bra{4}\rho_4\ket{2} = +i.
\end{align}
The analysis above shows that, assuming that the key transformations
are implemented perfectly, and that dissipative losses and experimental imperfections are negligible, double-quantum coherence is generated with maximum amplitude.

\subsection{GeoDQ: Implementation}
The GeoDQ procedure requires an implementation of the key element shown as $R_z^{13}(\pi)$ in Figure~\ref{fig:DQexc-schemes}(b). This element implements a cyclic trajectory in the $\{\ket{1},\ket{3}\}$ zero-quantum subspace. The trajectory subtends a solid angle about the origin of the subspace, which induces a geometric phase for the rotated quantum states. The geometry of the trajectory is chosen such that the propagator corresponds to a $\pi$ rotation about the $z$-axis of this subspace, corresponding to the operator $R_z^{13}(\pi)$, as depicted in Figure~\ref{fig:DQ-excitation-GeoDQ}. Some concrete implementations of this element, which involve spin echo trains of strong $\pi$ pulses separated by delays, are given in refs.~\cite{bengs_aharonov_2023,heramun_singlet_2025}. The experimental implementation used for the current work is specified in Section~\ref{sec:Results}.

\begin{figure}[tb]
\centering
\includegraphics[width=0.85\columnwidth]{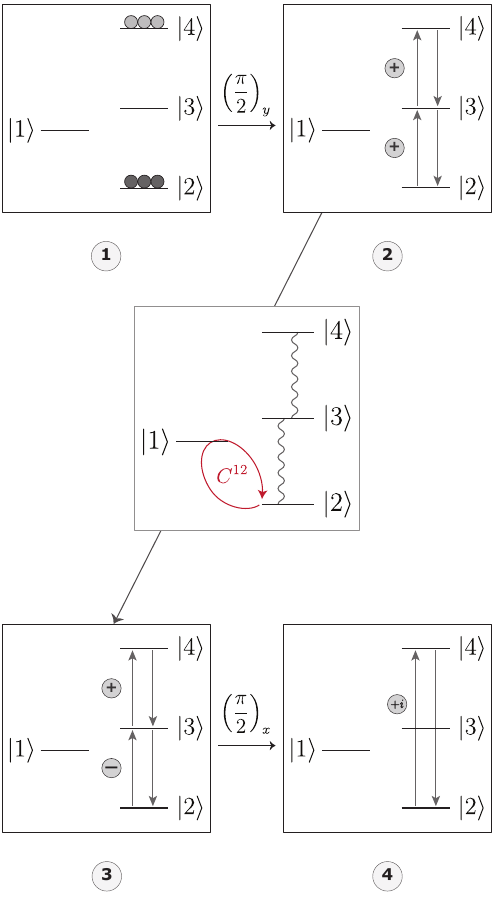}
\caption{
Spinor double-quantum excitation. An ensemble of spin-1/2 pairs is at thermal equilibrium in a strong magnetic field (pane 1). In-phase single-quantum triplet-triplet coherences are generated by a $\pi/2$ $y$-pulse (pane 2). This is converted into a \emph{double-quantum precursor state} of antiphase triplet-triplet coherences (pane 3). A final $\pi/2$ $x$-pulse generates double-quantum coherence between the states $\ket{T_{\pm1}}$ (pane 4). This method uses a rotation through $2\pi$ in the single-quantum subspace spanned by the singlet state $\Szero=\ket{1}$ and the outer triplet state $\Tpos=\ket{2}$. The rotation corresponds to the single-transition cycle operator $C^{12}$ (Equation~\ref{eq:C12}). 
}
\label{fig:DQ-excitation-Spinor-DQ}
\end{figure} 
\section{Spinor double-quantum excitation}
\label{sec:Spinor-DQ}

\subsection{General principles}

\subsubsection{Single-quantum excitation}
The first two steps in the Spinor-DQ excitation scheme are the same as for geometric DQ excitation. In-phase single-quantum coherence is excited by applying a $\pi/2$ pulse to a system in thermal equilibrium. This leads to the state $\rho_2$ given in equation~\ref{eq:rho2}, and depicted graphically in the second pane of figure~\ref{fig:DQ-excitation-Spinor-DQ}. 

\subsubsection{Preparation of the double-quantum precursor}
Spinor-DQ prepares the double-quantum precursor state by a different method to that used in GeoDQ. Instead of implementing a $\pi$ rotation in the $\{\ket{1},\ket{3}\}$ subspace, Spinor-DQ implements a $2\pi$ rotation in an outer singlet-triplet subspace. There are two such subspaces; the lower one, spanned by the states $
\{\ket{S_0},\ket{T_{+1}}\}=
\{\ket{1},\ket{2}\}$, and the upper one, spanned by the states $\{\ket{S_0},\ket{T_{-1}}\}=\{\ket{1},\ket{4}\}$.  
The following discussion assumes that a cycle is implemented in the $\{\ket{1},\ket{2}\}$ subspace, as shown by the central pane in Figure~\ref{fig:DQ-excitation-Spinor-DQ}. 

The relevant transformation of the spin density operator is as follows:
\begin{align}
\label{eq:rho3spinorDQ}
\rho_3^\mathrm{spinor} &=
 C^{12}\,\rho_2\, C^{12\dagger}
 \nonumber\\
 &= 2^{1/2} C^{12}
 \left(
 I_{x}^{23}+I_{x}^{34}
 \right)
 C^{12\dagger}
 \nonumber\\
 &= 2^{1/2}
 \left(
 -I_{x}^{23}+I_{x}^{34}
 \right)
 \nonumber\\
&= 2^{-1/2}\,\big\{
-\ket{2}\bra{3}-\ket{3}\bra{2}
 \nonumber\\
&\qquad\qquad\qquad
+\ket{3}\bra{4}+\ket{4}\bra{3}
\big\}.
\end{align}
This transformation follows from the properties of the single-transition cycle operator in Equation~\ref{eq:C12}:
\begin{align}
C^{12}\,\ket{2}&=-\ket{2},
\nonumber\\
C^{12}\,\ket{3}&=+\ket{3},
\nonumber\\
C^{12}\,\ket{4}&=+\ket{4},
\end{align}
The change of sign of $\ket{2}$ by the cycle operator $C^{12}$ is a manifestation of spinor behaviour. A graphical representation of $\rho_3^\mathrm{spinor}$ is shown in pane 3 of Figure~\ref{fig:DQ-excitation-Spinor-DQ}. 

The density operator $\rho_3^\mathrm{spinor}$ may be written in terms of Cartesian product operators~\cite{EBW_1987,nakai_inadequate_1993} as follows:
\begin{equation}
\label{eq:rho3spinorCart}
\rho_3^\mathrm{spinor} = -2I_{1x}I_{2z}-2I_{1z}I_{2x}.
\end{equation}

\subsubsection{Double-quantum excitation}
The double-quantum precursor state $\rho_3^\mathrm{spinor}$ is converted into double-quantum coherence by a final $\pi/2$ pulse. 
In the Spinor-DQ procedure, the pulse phase is $0$, leading to the following result:
\begin{align}
\label{eq:rho4spinor}
\rho_4
&=R_x(\pi/2)\,\rho_3^\mathrm{spinor}\, R_x(\pi/2)^\dagger
 \nonumber\\
 &=-R_x(\pi/2)\,
 (2I_{1x}I_{2z}+2I_{1z}I_{2x})
  R_x(\pi/2)^\dagger
\nonumber\\
&=
 2I_{1x}I_{2y}+2I_{1y}I_{2x}.
\end{align}
This is exactly the same final state as for geometric double-quantum excitation (equation~\ref{eq:rho4}). A graphical representation is shown in pane 4 of Figure~\ref{fig:DQ-excitation-Spinor-DQ}. 

Hence, in the absence of imperfections and dissipative losses, both GeoDQ and Spinor-DQ generate double-quantum coherence with maximum amplitude.

\subsection{Spinor DQ: PulsePol/Symmetry-Based Implementation}

The Spinor-DQ procedure requires an implementation of the key element shown as $C^{12}(\pi)$ in Figure~\ref{fig:DQexc-schemes}(c). This element implements a $2\pi$ rotation in the $\{\ket{1},\ket{2}\}$ single-quantum subspace, as depicted in the central pane of Figure~\ref{fig:DQ-excitation-Spinor-DQ}. The PulsePol/symmetry-based implementation of Spinor-DQ is now discussed. 

\subsubsection{PulsePol as a Symmetry-Based Pulse Sequence}
\label{sec:PulsePol-DQ}
PulsePol is a repeating sequence of six strong pulses and four delays~\cite{schwartz_robust_2018}, as follows:
\begin{multline}
\label{eq:PulsePol}
\mathrm{PulsePol}=
90_{90} -\tau_2- 180_0 -\tau_2- 90_{90}-
\\
90_{180} -\tau_2- 180_{90} -\tau_2- 90_{180}
\qquad
\end{multline}

As shown in Ref.~\cite{sabba_symmetrybased_2022}, PulsePol is closely related to the symmetry-based sequence denoted here as $R4_3^1$, which has the following explicit form:
\begin{multline}
\label{eq:R431}
R4_3^1 = 
\big[\,
90_{135} -\tau_2- 180_{45} -\tau_2- 90_{135} -
\\
90_{-135} -\tau_2- 180_{135} -\tau_2- 90_{-135}
\,\big]^2
\end{multline}
where the superscript denotes 2 repetitions. When applied to spin-1/2 pairs in solution NMR, the sequence in Equation~\ref{eq:R431} conforms to $R4_3^1$ symmetry, if the pulse sequence intervals $\tau_2$ are given, in the strong-pulse limit, by: 
\begin{equation}
\label{eq:tau2PulsePol}
    \tau_2 = \tfrac38 \tau_J.
\end{equation}
where the $J$-coupling period is given by:
\begin{equation}
\label{eq:tauJ}
    \tau_J = \frac{2\pi}{|\omega_J|}=\vert J\vert^{-1}.
\end{equation}

In general, pulse sequences with the symmetry designation $RN_n^\nu$ are constructed by starting with a \enquote{basic element}, denoted $R^0$, and applying a set of transformations which depend on the values of three integers, called symmetry numbers, and denoted $N$, $n$ and $\nu$. For the implementation of $R4_3^1$ in Equation~\ref{eq:R431}, the basic element is as follows: 
\begin{equation}
\label{eq:R0}
    R^0 = 90_{90}-\tau_2 - 180_{0} -\tau_2 - 90_{90} \ ,
\end{equation}
This is a standard composite $\pi$ pulse~\cite{levitt_nmr_1979}, but with delays inserted between the pulses. For $R4_3^1$, the symmetry numbers are $\{N,n,\nu\}=\{4,3,1\}$. This implies that the complete $R4_3^1$ sequence consists of four $R$-elements, spanning the same duration as three $J$-periods, with the phases of the $R$-elements alternated between the values $\pm\phi$, where $\phi=\pi\nu/N=\pi/4$ in the case of $R4_3^1$. The total duration of the $R4_3^1$ sequence is given by:
\begin{equation}
\label{eq:TR4-matches-3tauJ}
    T_{R4} = 3\tau_J.
\end{equation} 
Other combinations of symmetry numbers are also possible, as specified in Table $1$ of Ref.~\cite{sabba_symmetrybased_2022}. This rational, symmetry-based, construction procedure leads to well-defined selection rules for the average Hamiltonian $\overline{H}^{(1)}$, which governs the dynamical properties of the pulse sequence, within certain approximations~\cite{levitt_symmetrybased_2007,levitt_symmetry_2008}. 

The $R4_3^1$ and PulsePol sequences are related as follows:
\begin{equation}
\label{eq:R-PulsePol-correspondence}
    R4_3^1 = \left[\left( \mathrm{PulsePol} \right)_{\phi=\pi/4}\right]^2 \ ,
\end{equation}
implying that $R4_3^1$ is a two-fold repetition of a phase-shifted PulsePol sequence. The following discussion considers the $R4_3^1$ sequence in Equation~\ref{eq:R431}, rather than the closely-related PulsePol sequence in Equation~\ref{eq:PulsePol}, since this allows the direct application of symmetry-based pulse sequence theory~\cite{lee_efficient_1995,carravetta_symmetry_2000,levitt_symmetrybased_2007,levitt_symmetry_2008}.  Furthermore, the phase behaviour of $R4_3^1$ is more convenient than that of PulsePol, as discussed below. 

The sequence in Equation~\ref{eq:R431} uses a variant construction procedure, in which two basic elements are alternated in a \enquote{riffling} scheme, in order to improve robustness~\cite{sabba_symmetrybased_2022}. Nevertheless, in the limit of short rf pulses, the symmetry properties and selection rules for $R4_3^1$ symmetry still apply. 

\subsubsection{Cycle Propagator}
The PulsePol/Symmetry-Based implementation of Spinor-DQ excitation uses the sequence shown in Figure~\ref{fig:DQexc-schemes}(c), with a central element $C^{12}$ given by $m$ repetitions of a $R4_3^1$ sequence (and hence $2m$ repetitions of a phase-shifted PulsePol sequence):
\begin{equation}
\label{eq:C12-R431}
    C^{12} = \left[R4_3^1 \right]^m .
\end{equation}
The total duration of the PulsePol/Symmetry-Based double-quantum excitation sequence is therefore given by:
\begin{equation}
    T = m T_{R4} = 3 m \tau_J
    = 24 m \tau_2.
\end{equation}
The choice of repetition number $m$ is discussed below.  
As described in Ref.~\cite{sabba_symmetrybased_2022}, the theory of $R4_3^1$ in the current context proceeds by describing the chemical shift Hamiltonians $H_\Delta$ and $H_\Sigma$ in the interaction frame of the $J$-coupling and rf fields:
\begin{align}
\label{eq:Htilde-terms}
\tilde{H}_\Delta &=
U_\mathrm{rf}(t)^\dagger
U_J(t)^\dagger
H_\Delta
U_J(t)U_\mathrm{rf}(t),
\nonumber\\
\tilde{H}_\Sigma &=
U_\mathrm{rf}(t)^\dagger
U_J(t)^\dagger
H_\Sigma
U_J(t)U_\mathrm{rf}(t),
\end{align}
where $U_\mathrm{rf}(t)$ is the propagator under the rf fields of the pulse sequence. The propagator under the $J$-coupling Hamiltonian is defined as follows:
\begin{equation}
\label{eq:UJ}
    U_J(t)=\exp\{-i H_J t\}.
\end{equation}
The propagator for the spin system, at any time point $t$, may be approximated as follows:
\begin{equation}
\label{eq:U(t)-R431}
U(t) \simeq
U_J(t)U_\mathrm{rf}(t)
\times
\exp\{-i \overline{H}^{(1)} t \},
\end{equation}
where $\overline{H}^{(1)}$ is the average Hamiltonian~\cite{haeberlen_coherent_1968, haeberlen_high_1976, hohwy_elimination_1997} of the chemical shift terms over a single $R4_3^1$ sequence:
\begin{equation}
\overline{H}^{(1)} =
\overline{H}_\Delta^{(1)}+
\overline{H}_\Sigma^{(1)},
\end{equation}
with:
\begin{equation}
\overline{H}_\Delta^{(1)} = 
T_{R4}^{-1}
\int_0^{T_{R4}}
\tilde{H}_\Delta(t)
\ dt,
\end{equation}
and similarly for $\overline{H}_\Sigma^{(1)}$. The numbering convention used here for the average Hamiltonian terms is that of Hohwy \textit{et al.}~\cite{hohwy_elimination_1997}, superseding an earlier convention~\cite{haeberlen_coherent_1968,haeberlen_high_1976}.

The selection rules associated with the symmetry number combinations given in Table 1 of Ref.~\cite{sabba_symmetrybased_2022} lead to an average Hamiltonian with the following form:
\begin{equation}
    \overline{H}^{(1)} 
    =
     \overline{H}^{(1)}_{1111}
     +\overline{H}^{(1)}_{1\,-11\,-1},
\end{equation}
where:
\begin{align}
    \overline{H}^{(1)}_{1111}
    &=
    \tfrac12\kappa_{1111}\wD \ket{2}\bra{1},
\nonumber\\
    \overline{H}^{(1)}_{1\,-11\,-1}
    &=
    -\tfrac12\kappa_{1\,-11\,-1}\wD \ket{1}\bra{2}.
\end{align}

Note that the $R4_3^1$ symmetry causes all terms derived from $H_\Sigma$ to vanish. 
The scaling factors $\kappa_{1\,\pm1 1\,\pm1}$ depend on the symmetry numbers $\{N, n, \nu\}$ of the pulse sequence and the details of the basic element $R^0$. General expressions are given in Ref.~\cite{sabba_symmetrybased_2022}.
In the limit of infinitesimally short pulses, the scaling factors $\kappa_{1\,\pm1 1\,\pm1}$ for the basic element of Equation~\ref{eq:R0} are given by:
\begin{align}
\label{eq:kappa}
\kappa_{1111} &= -\kappa_{1\,-11\,-1}
=(-1)^p \frac{\sqrt{2}N}{\pi n}\sin^2\!\left(\frac{\pi n}{2 N}\right) , 
\nonumber\\
p &= \frac{n-\nu}{N} -\tfrac{1}{2}; \quad p \in \mathbb{Z},
\end{align}
where $\mathbb{Z}$ is the set of all integers.

In the case of the $R4_3^1$ sequence, $\{N,n,\nu\}=\{4,3,1\}$, the scaling factors evaluate to:
\begin{align}
\label{eq:kappa}
\kappa_{1111} = -\kappa_{1\,-11\,-1}
=\frac{2(1+\sqrt2)}{3\pi}\,
\simeq 0.512\ ,
\end{align}
again assuming infinitesimally short pulses. 

The first-order average Hamiltonian for the $R4_3^1$ sequence is therefore given by:
\begin{align}
\label{eq:avHamR431}
\overline{H}^{(1)} =
\omega^{12}_\mathrm{nut} I_x^{12},
\end{align}
where the single-transition nutation frequency is proportional to the chemical shift difference frequency $\wD$:
\begin{align}
\label{eq:w12nut}
\omega^{12}_\mathrm{nut} =
\kappa_{1111} \wD \simeq 0.512\,\wD.
\end{align}

The selection rules imposed by $R4_3^1$ symmetry ensure that the average Hamiltonian is selective for the $\{\ket{1},\ket{2}\}$ transition. The real values of the scaling factors $\kappa_{1\,\pm1 1\,\pm1}$ indicate that the $R4_3^1$ sequence generates a rotation about the $x$-axis of the $\{\ket{1},\ket{2}\}$ subspace. In contrast, the effective rotation axis of the PulsePol sequence (Equation~\ref{eq:PulsePol}) is rotated away from the $x$-axis by $45^{\circ}$. 

From Equation~\ref{eq:U(t)-R431}, the propagator for a single $R4_3^1$ sequence is  given by:
\begin{align}
\label{eq:UR431}
U(R4_3^1) &\simeq 
U_J(T_{R4})
U_\mathrm{rf}(R4_3^1)
R_x^{12}(\omega^{12}_\mathrm{nut} T_{R4}),
\end{align}
where the single-transition rotation angle is:
\begin{equation}
\omega^{12}_\mathrm{nut} T_{R4} 
=8\omega^{12}_\mathrm{nut}\tau_2,
\end{equation}
assuming strong pulses of infinitesimal duration, separated by intervals $\tau_2$, as shown by Equation~\ref{eq:R431}. The single-transition rotation operator $R_x^{12}$ is defined in Equation~\ref{eq:Rrs}.

The propagators $U_J$
and $U_\mathrm{rf}$ are both unity operators, when taken over a complete $R4_3^1$ sequence. Hence the propagator for a complete $R4_3^1$ sequence is given by a rotation around the $x$-axis of the $\{\ket{1},\ket{2}\}$ subspace:
\begin{align}
\label{eq:UR431}
U(R4_3^1) &\simeq 
R_x^{12}(\omega^{12}_\mathrm{nut} T_{R4}).
\end{align} 
If the $R4_3^1$ sequence is repeated $m$ times, the total duration of the entire sequence is given by:
\begin{equation}
    T = m T_{R4} = 3 m \tau_J
    = 24 m \tau_2.
\end{equation}
The overall spin propagator is given by:
\begin{equation}
    U(\left[R4_3^1\right]^m) \simeq 
    R_x^{12}(\beta^{12}),
\end{equation}
where the total rotation angle in the $\{\ket{1},\ket{2}\}$ subspace  is: 
\begin{equation}
\label{eq:beta12}
\beta^{12} = \omega^{12}_\mathrm{nut} T.
\end{equation}
The transformation of the density operator $\rho_2$, as given by Equation~\ref{eq:rho2}, takes the following form:
\begin{align}
\label{eq:rho3R431}
\rho_3(T) &\simeq
 R_x^{12}(\beta^{12})\,\rho_2\, 
 R_x^{12}(\beta^{12})^\dagger
 \nonumber\\
 &{\!\!\!\!\!\!\!\!\!\!}{\!\!\!\!\!\!\!\!\!\!}
 = 2^{1/2}  R_x^{12}(\beta^{12})
 \left(
 I_{x}^{23}+I_{x}^{34}
 \right)
R_x^{12}(\beta^{12})^\dagger
 \nonumber\\
 &{\!\!\!\!\!\!\!\!\!\!}{\!\!\!\!\!\!\!\!\!\!}
 = 2^{1/2}
 \big(
 I_{x}^{23}\cos(\tfrac12 \beta^{12})
+I_{y}^{13}\sin(\tfrac12 \beta^{12})
 +I_{x}^{34}
 \big).
 \nonumber\\
\end{align}
The last equation follows from the commutation relationships~\cite{vega_fictitious_1978, wokaun_selective_1977}:
\begin{align}
[ I_x^{12}, I_x^{23}] &= \tfrac{1}{2}i I_y^{13},
\nonumber\\
[ I_x^{12}, I_y^{13}] &= -\tfrac{1}{2}i I_x^{23},
\nonumber\\
[ I_x^{12}, I_x^{34}] &= 0.
\end{align}
Note the $4\pi$-periodicity with respect to $\beta^{12}$ in Equation~\ref{eq:rho3R431}. This is a signature of spinor behaviour. 

\subsubsection{Double-Quantum Excitation}
The total rotation angle $\beta^{12}$ may be set to $\approx 2\pi$ by choosing the repetition number $m$ as follows:
\begin{equation}
m = 
\mathrm{round}\!\left(
\frac{2\pi}{8|\kappa_{1111}\,\wD|\tau_{2}  } 
\right)
\simeq 
\mathrm{round}\!\left(
0.651 \frac{J}{\Delta}
\right).
\end{equation}
With this choice of repetition number $m$, the propagator for the repeated $R4_3^1$ sequences approximates a cycle:
\begin{equation}
   U(\left[R4_3^1\right]^m) \simeq 
   R_x^{12}(2\pi) =
   C^{12}. 
\end{equation}
Under these conditions, the transformations of  Equations~\ref{eq:rho3spinorDQ} and \ref{eq:rho4spinor} follow to a good approximation. Double-quantum coherence is generated with large amplitude by the final $(\pi/2)_x$ pulse, as sketched in Figure~\ref{fig:DQ-excitation-Spinor-DQ}. 

Note that the sequence duration $T$ required to achieve Spinor-DQ excitation is twice that needed to achieve the generation of singlet order\cite{sabba_symmetrybased_2022}. This because Spinor-DQ excitation requires a $2\pi$ rotation in the $\{\ket{1},\ket{2}\}$ subspace, while the generation of singlet order by PulsePol only requires a $\pi$ rotation.

In general, the spin density operator at the end of the complete sequence has the following form:
\begin{align}
\rho_4(T)&=
R_x(\pi/2)\rho_3(T)R_x^\dagger(\pi/2)
\nonumber\\
  &\simeq 
2 I_y^{24}
\sin^2(\tfrac{1}{4}\beta^{12})
+\ldots  
\end{align}
which shows the excitation of double-quantum coherence as well as other orthogonal terms, in the general case (omitted for simplicity).

In general, the double-quantum amplitude is given in terms of the total pulse sequence duration $T$ as follows:
\begin{equation}
\label{eq:aDQpulsepol}
 a_\mathrm{DQ}^{\mathrm{PulsePol}}(T)  \simeq -i \sin^2\!\left(\tfrac{1}{4} \omega_{\mathrm{nut}}^{12} T\right),    
\end{equation}
where the effective nutation frequency $\omega_{\mathrm{nut}}^{12}$ of the $\{\ket{1},\ket{2}\}$ transition is given by Equation~\ref{eq:w12nut}. 

\begin{figure}[tb] 
\includegraphics[width=0.9\columnwidth]{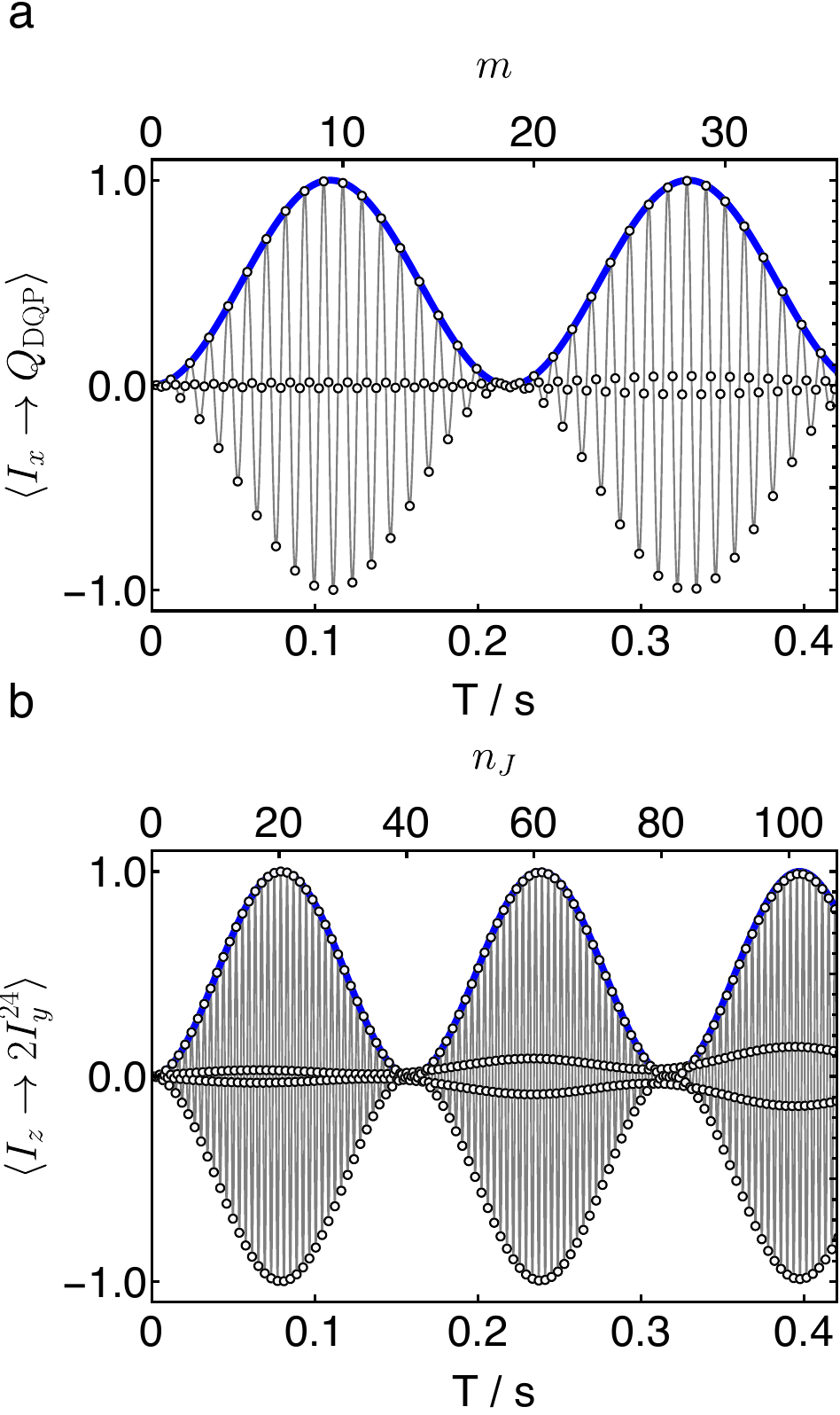} 
\caption{
Excitation trajectories for spinor double-quantum excitation schemes applied to isolated 2-spin-1/2 systems.
(a): A PulsePol $R4_3^1$ sequence is applied to transverse magnetization $I_x$, converting it to the double-quantum precursor state $Q_{\rm{DQP}}$. The plotted amplitude $<I_x\to Q_\mathrm{DQP}>$ is defined in equations~\ref{eq:TransformationAmplitude} and \ref{eq:Q-DQP}. 
(b): A SLIC pulse, as shown in Figure~\ref{fig:SLIC-DQ-sequences}(a), is applied to longitudinal magnetization $I_z$, converting it to the double-quantum state $2I_{y}^{24}$. 
Solid blue lines: predictions from average Hamiltonian theory (Equations~\ref{eq:aDQpulsepol} and \ref{eq:aDQSLICa}).
Solid gray lines: predictions from average Hamiltonian theory multiplied by functions $\cos(\omega_{J} t)$ (for SLIC-DQ) or $\cos(\tfrac13\omega_{J} t)$ (for PulsePol-DQ).
Open circles: simulated points. 
All simulations use $J = 255.94$~Hz and $\Delta=17.8$~Hz. 
}
\label{fig:ExcitationTrajectories}
\end{figure} 

A numerical simulation of PulsePol/symmetry-based double-quantum excitation is shown in Figure~\ref{fig:ExcitationTrajectories}(a). This shows the trajectory of the transformation amplitude $\langle I_x \overset{U}{\to} Q_\mathrm{DQP}\rangle$ (see Equation~\ref{eq:TransformationAmplitude}) for conversion of the state $\rho_2$ (Equation~\ref{eq:rho2}) into the desired double-quantum precursor state, defined as follows:
\begin{equation}
    \label{eq:Q-DQP}
    Q_\mathrm{DQP} 
    = 2^{1/2}(I_x^{34}-I_x^{23}).
\end{equation}

Since the double-quantum precursor operator $Q_\mathrm{DQP}$ is completely converted into double-quantum coherence by the final $\pi/2$ pulse, the quantity $\langle I_x \overset{U}{\to} Q_\mathrm{DQP}\rangle$, defined in equation~\ref{eq:TransformationAmplitude}, provides the double-quantum excitation amplitude of the full pulse sequence, including the last $\pi/2$ pulse.

Figure~\ref{fig:ExcitationTrajectories}(a) also shows a comparison with the analytical result of Equation~\ref{eq:aDQpulsepol}. Agreement is good for an integer number of complete $R4_3^1$ cycles, as expected from average Hamiltonian theory. 

\subsubsection{Double-Quantum Filtration}
The double-quantum filtration amplitude by the PulsePol/Symmetry-Based scheme is given, in the absence of relaxation, by the square magnitude of Equation~\ref{eq:aDQpulsepol}:
\begin{equation}
\label{eq:aDQFpulsepol}
 a_\mathrm{DQF}^{\mathrm{PulsePol}}(T)  \simeq 
 \sin^4 
 \!\left(\tfrac{1}{4} \omega_{\mathrm{nut}}^{12} T\right).    
\end{equation}
This function is compared with numerical simulations of the double-quantum filtering amplitude, and experimental results, in Figure~\ref{fig:DQtrajectories}(b).

The requirement to complete an integer number of $R4_3^1$ cycles can be overly restrictive on the excitation time $T$. In practice, it is possible to truncate the double-quantum excitation before an integer number of $R4_3^1$ cycles is completed, providing that precautions are taken. This topic is discussed further in the Supporting Information.

\begin{figure}[tbh]
\centering
\includegraphics[width=\columnwidth]{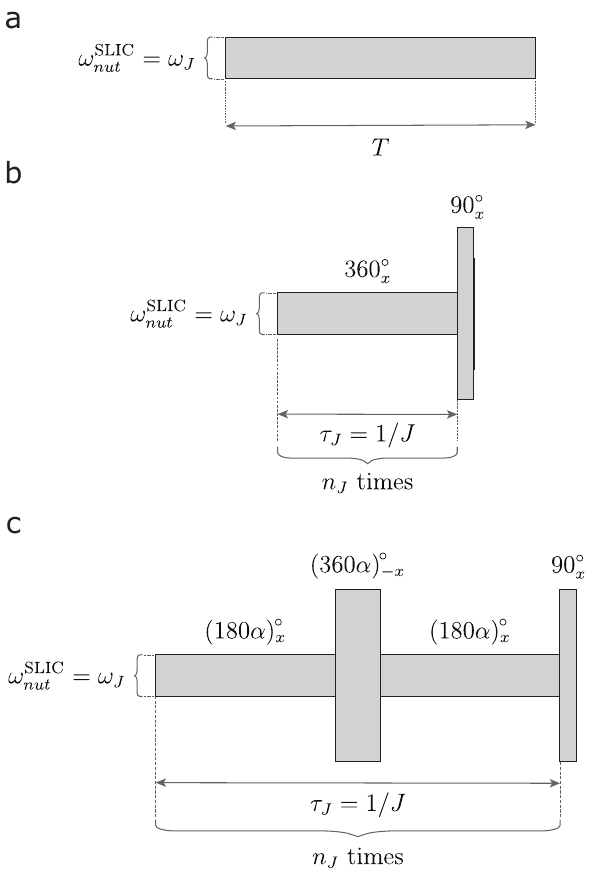}
\caption{
SLIC-DQ sequences.
({a}) Double-quantum excitation is induced by a weak rf field satisfying the SLIC condition 
$\omega_{\rm{nut}}^{\rm{SLIC}} = \omega_J$ for a total duration $T$. 
({b}) A strong $90\degree$ pulse is appended to the SLIC pulse, without a change of phase. The SLIC pulse is subdivided into a series of $n_J$ $360\degree_x$ pulses to ensure stroboscopic evolution. Optimal double-quantum excitation is achieved when the repeat number $n_J$ satisfies equation~\ref{eq:nJSLIC}.
({c}) Double-quantum excitation scheme using cSLIC, with compensation against rf amplitude errors. A strong central pulse with a $180\degree$ phase shift is inserted at the centre point of the SLIC pulse. 
The flip angle multiplier $\alpha$, with $\alpha\lesssim1$, is defined in equation~\ref{eq:alpha factor} and takes into account the finite duration of the central compensation pulse.  
For an infinitely short central pulse, $\alpha=1$, and the central pulse has a flip angle of $360\degree$. 
}
\label{fig:SLIC-DQ-sequences}
\end{figure} 

\subsection{Spinor DQ: SLIC implementation}
\label{sec:SLIC-DQ}

This section discusses the principles of double-quantum excitation in near-equivalent spin-1/2 pairs by SLIC and its variants. We call this approach SLIC-DQ. As explained below, SLIC-DQ is an example of a Spinor-DQ excitation scheme, albeit disguised by an implicit frame transformation.

\subsubsection{SLIC-DQ: Single Pulse Implementation}

The simplest SLIC implementation of Spinor-DQ excitation is shown in Figure~\ref{fig:SLIC-DQ-sequences}(a). This is a single, unmodulated, resonant radiofrequency pulse with phase $\phi=0$ and duration $T$.
The amplitude of the pulse is such that the nutation frequency under the pulse matches the $J$-coupling between the members of the spin-pair:
\begin{equation}
\label{eq:SLICcondition}
\mathrm{SLIC\ condition:}\qquad
\wnut =\wJ. 
\end{equation}
This spin-lock-induced crossing (SLIC) condition causes a level anticrossing between a pair of rotating-frame energy levels, in the presence of the resonant radiofrequency field~\cite{devience_preparation_2013a}. SLIC pulses are useful devices for manipulating nuclear singlet order~\cite{devience_preparation_2013a,
devience_nuclear_2013,
elliott_longlived_2016,
elliott_longlived_2016a,
eills_singlet_2017,
sheberstov_cis_2018,
elliott_nuclear_2019,
sheberstov_excitation_2019,
sheberstov_generating_2019,
devience_manipulating_2020,
korenchan_31p_2022,
sonnefeld_polychromatic_2022,
sonnefeld_longlived_2022,
razanahoera_hyperpolarization_2024,
sheberstov_collective_2024,
wiame_longlived_2025,
mandzhieva_zerofield_2025,
mcbride_scalable_2025}, 
for enhancing the contrast in MRI~\cite{devience_nuclear_2013}, for estimating $J$-couplings~\cite{devience_homonuclear_2021}, and in parahydrogen-enhanced NMR~\cite{eills_singlet_2017}. Here, we show that SLIC may also be used to excite homonuclear double-quantum coherence, albeit with a high degree of sensitivity to variations in the radiofrequency field amplitude. 

Note that the single SLIC pulse in Figure~\ref{fig:SLIC-DQ-sequences}(a) is not preceded by a $90\degree$ pulse, and not followed by one either. The single unadorned SLIC pulse has phase $\phi=0$ and is applied directly to thermal equilibrium magnetization, and generates double-quantum coherence without further manipulations, under the conditions discussed below. 

Although the SLIC phenomenon has been widely exploited in NMR, especially for the preparation and study of long-lived states~\cite{devience_preparation_2013a,
devience_nuclear_2013,
elliott_longlived_2016,
elliott_longlived_2016a,
eills_singlet_2017,
sheberstov_cis_2018,
elliott_nuclear_2019,
sheberstov_excitation_2019,
sheberstov_generating_2019,
devience_manipulating_2020,
korenchan_31p_2022,
sonnefeld_polychromatic_2022,
sonnefeld_longlived_2022,
razanahoera_hyperpolarization_2024,
sheberstov_collective_2024,
wiame_longlived_2025,
mandzhieva_zerofield_2025,
mcbride_scalable_2025},
the spin dynamics occurring during the SLIC pulse has mainly been discussed in terms of spin population transfers. The use of SLIC for directly exciting double-quantum coherences, of relevance here, requires a more detailed theoretical treatment.  

To avoid potential confusion, the excitation of double-quantum coherence by application of a SLIC-matched radiofrequency field does \emph{not} exploit the double-quantum level-anticrossing conditions which have been identified in systems of more than 2 coupled spins~\cite{sonnefeld_polychromatic_2022}. The methods under discussion here achieve double-quantum excitation by SLIC pulses which conform to the conventional \emph{single-quantum} level anti-crossing condition of Equation~\ref{eq:SLICcondition}.

The spin Hamiltonian in the presence of the SLIC pulse is given by:
\begin{equation}
\label{eq:HSLIC}
  H = H_\Sigma + H_\Delta + H_J 
  + \wnut I_x.  
\end{equation}
The individual Hamiltonian terms are specified in Equation~\ref{eq:H0terms}.

The SLIC pulse is applied to a thermal-equilibrium spin system in high magnetic field, described by the density operator $\rho_1=I_z$ in Equation~\ref{eq:rho1}. If the SLIC pulse has duration $T$, the final density operator is given by:
\begin{equation}
\label{eq:rhofSLIC}
\rho_f(T) =
U(T)
\,I_z\,
U^\dagger(T)
\ ,
\end{equation}
where the SLIC propagator is:
\begin{equation}
  U(T)
  =
  \exp\{-i H T\},
\end{equation}
and relaxation effects have been omitted.

The analysis is facilitated by transforming the Hamiltonian into a periodic interaction frame, defined by the following unitary transformation of spin operators:
\begin{equation}
\tilde{Q}=
W(t)^\dagger\,Q\,W(t),
\end{equation}
where the time-dependent transformation operator is defined here as follows:
\begin{equation}
\label{eq:W(t)}
W(t) = R_y(-\pi/2)U_J(t)R_z(-\wJ t),
\end{equation}
where $U_J$ is given by Equation~\ref{eq:UJ} and $0\leq t \leq T$. This frame is periodic, with a period $\tau_J=|J|^{-1}$, since the following property holds for all time-independent operators $Q$ and times $t$:
\begin{equation}
\label{eq:periodicQtilde}
\tilde{Q}(t+k\tau_J) =
\tilde{Q}(t)
\ ;\quad k\in\mathbb{Z}.
\end{equation}

The equation of motion of the spin density operator in the interaction frame is given by:
\begin{equation}
\frac{d}{dt}\tilde{\rho}
=
-i \big[ 
\tilde{H}, \tilde{\rho}
\big],
\end{equation}
where the interaction-frame spin Hamiltonian is given by:
\begin{align}
\label{eq:Htilde}
\tilde{H}(t) 
&=
W(t)^\dagger\,H\,W(t)
-i\,W(t)^\dagger
\Big(\frac{d}{dt}W(t)\Big).
\end{align}
The second term in Equation~\ref{eq:Htilde} corrects for the non-inertial motion of the frame, playing a similar role to the Coriolis force in classical mechanics~\cite{cohen-tannoudji_quantum_1977,eden_zeeman_2015,chavez_interaction_2022}.

These equations may be combined to obtain the
interaction frame Hamiltonian:
\begin{align}
\tilde{H}(t) 
&=
W(t)^\dagger
H
W(t)
-H_J + \wJ I_z
\nonumber\\
&=
\tilde{H}_\Delta(t) +
\tilde{H}_\Sigma(t) 
- (\wnut - \wJ) I_z
\nonumber\\
&=
\tilde{H}_\Delta(t) +
\tilde{H}_\Sigma(t) 
- \epsilon_\mathrm{rf}\, \wJ I_z,
\end{align}
where the fractional rf amplitude mismatch is defined as follows:
\begin{equation}
\label{eq:erf}
\epsilon_\mathrm{rf} = \frac{\wnut - \wJ}{\wJ}.
\end{equation}
The fractional rf mismatch $\epsilon_\mathrm{rf}$ goes to zero for exact SLIC match (eq.~\ref{eq:SLICcondition}). 

The interaction-frame chemical shift terms are given by:
\begin{align}
\tilde{H}_\Delta(t) 
&=
\tfrac12\wD 
\,W(t)^\dagger\,
(I_{1z}-I_{2z})\,
W(t),
\nonumber\\
\tilde{H}_\Sigma(t) 
&=
\tfrac12\omega_\Sigma 
\,W(t)^\dagger\,
(I_{1z}+I_{2z})\,
W(t).
\end{align}

The interaction-frame Hamiltonian terms may be written in terms of single-transition operators, as follows:
\begin{align}
\label{eq:HDtilde+HStilde}
\tilde{H}_\Delta(t) 
&=
2^{-1/2}\wD\big(
-I_x^{12}
+I_x^{14}\cos(2\wJ t)
+I_y^{14}\sin(2\wJ t)
\big),
\nonumber\\
\tilde{H}_\Sigma(t) 
&=
\tfrac{1}{2}\omega_\Sigma\big(
I_x\cos(\wJ t)
+I_y\sin(\wJ t)
\big).
\end{align}
In the case that the chemical shift frequencies $\wD$ and $\omega_\Sigma$
are small compared to $\wJ$, and also $|\epsilon_\mathrm{rf}|\ll1$, the periodicity and cyclicity of the interaction frame (Equation~\ref{eq:periodicQtilde}) may be exploited to estimate the propagator for the spin system over the total interval $T$ from the average of the interaction frame Hamiltonian over one period:
\begin{equation}
\label{eq:USLIC}
U(T) \simeq
W(T)
\exp\{-i \overline{H}^{(1)} T\}
W(0)^\dagger.
\end{equation}
The average Hamiltonian for SLIC, in the interaction frame, is given by:
\begin{align}
\label{eq:HbarSLIC}
\overline{H}^{(1)}
&=\tau_J^{-1}
\int_0^{\tau_J}
\tilde{H}(t)\, dt
\nonumber\\
&=
- 2^{-1/2}
\,\wD I_x^{12}
- \epsilon_\mathrm{rf}\, \wJ I_z
\nonumber\\
&=
- \kappa_\mathrm{SLIC}
\,\wD I_x^{12}
- \epsilon_\mathrm{rf}\, \wJ I_z.
\end{align}
This shows that, on exact match, SLIC induces a rotation in the $\{\ket{1},\ket{2}\}$ subspace at a nutation frequency given by minus the shift frequency difference $\wD$ multiplied by a scaling factor, given by:
\begin{equation}
\label{eq:kappaSLIC}
\kappa_\mathrm{SLIC} = 2^{-1/2}.    
\end{equation}

The time-average over one period removes the oscillatory terms in Equation~\ref{eq:HDtilde+HStilde}. The approximation is valid for spin-1/2 pair systems in the near-equivalence limit, irradiated with a SLIC pulse close to resonance, and with an amplitude close to exact SLIC match. Higher-order corrections to Equation~\ref{eq:HbarSLIC} are discussed in Appendix~\ref{appx:SlicHigherOrderAvHam}.

The first term in Equation~\ref{eq:HbarSLIC} shows that, to a good approximation, a SLIC pulse generates a rotation in the $\{\ket{1},\ket{2}\}$ subspace of the 2-spin-1/2 system, in the interaction frame. This has the same form as that generated by a $R4_3^1$ sequence in the ordinary rotating frame (Equation~\ref{eq:avHamR431}). 

The second term in Equation~\ref{eq:HbarSLIC} represents the deviation of the rf field amplitude from the exact SLIC condition. Since these two terms do not commute, the rf mismatch term readily interferes with the desirable subspace rotation. 

Now consider double-quantum excitation by a single on-resonance SLIC pulse, as shown in Figure~\ref{fig:SLIC-DQ-sequences}(a), assuming exact SLIC match ($\epsilon_\mathrm{rf} =0$). The pulse is applied to thermal equilibrium magnetization along the magnetic field, as described by the density operator $\rho_1=I_z$ of Equation~\ref{eq:rho1}. 
Equation~\ref{eq:rhofSLIC} may be implemented in three steps. Application of the right-most operator in  Equation~\ref{eq:USLIC} gives:
\begin{align}
W(0)^\dagger 
\,I_z\,
W(0)
&=
R_y(\pi/2) 
\,I_z\,
R_y^\dagger(\pi/2)
= I_x
\nonumber\\
&=
2^{1/2}(I_x^{23}+I_x^{34}).
\end{align}
Note that this is identical to the density operator in pane 2 of the general scheme for Spinor-DQ excitation (Figure~\ref{fig:DQ-excitation-Spinor-DQ}). 

From Equation~\ref{eq:HbarSLIC}, the propagator under the average Hamiltonian $\overline{H}^{(1)}$ for an interval $T$, in the case $\epsilon_\mathrm{rf}=0$, is given by:
\begin{equation}
\exp\{-i \overline{H}^{(1)}
T\}
=
R_x^{12}(\beta^{12}),
\end{equation}
where the rotation angle in the $\{\ket{1},\ket{2}\}$ subspace is given for SLIC by:
\begin{equation}
\label{eq:beta12SLIC}
\beta^{12} =
- \kappa_\mathrm{SLIC}\wD T,
\end{equation}
with the SLIC scaling factor $\kappa_\mathrm{SLIC}=2^{-1/2}$.
Hence, the transformation under the average Hamiltonian is given by:
\begin{multline}
\label{eq:SLICstep2}
R_x^{12}(\beta^{12})
I_x
R_x^{12\,\dagger}(\beta^{12})
= \\
2^{1/2}
 \big(
 I_{x}^{23}\cos(\tfrac12 \beta^{12})
+I_{y}^{13}\sin(\tfrac12 \beta^{12})
 +I_{x}^{34}
 \big).
\end{multline}
This expression is $4\pi$-periodic in the angle $\beta^{12}$, a signature of spinor behaviour. 
This transformation is similar to that given by a $R4_3^1$ sequence (Equation~\ref{eq:rho3R431}).

The total rotation angle $\beta^{12}$ may be set to $\approx -2\pi$ by choosing the interval $\tau$ as follows:
\begin{equation}
\label{eq:tauSLICDQopt}
T 
\simeq
\mathrm{round}\!\left(
\frac{2\pi J}{|\kappa_\mathrm{SLIC}\,\wD |} 
\right)\tau_J
\simeq 
\mathrm{round}\!\left(\left|
\frac{\sqrt{2} J}{\Delta}
\right|\right)\tau_J.
\end{equation}
With this choice of $T$, the evolution under the SLIC average Hamiltonian approximates a cycle, the spinor behaviour is activated, and  
Equation~\ref{eq:SLICstep2} evaluates to:
\begin{equation}
R_x^{12}(\beta^{12})
I_x
R_x^{12\,\dagger}(\beta^{12})
\simeq
2^{1/2}
 \big(
- I_{x}^{23}
 +I_{x}^{34}
 \big).
\end{equation}
The opposite signs indicate the generation of the double-quantum precursor state (antiphase single-quantum coherence), which corresponds to panel 3 of Figure~\ref{fig:DQ-excitation-Spinor-DQ}.

In general, the spin density operator at the end of the SLIC pulse may be calculated by applying the operator $W(T)$ to the right-hand side of Equation~\ref{eq:SLICstep2}. This leads to 
\begin{align}
\rho_f(T) &=
-2 I_y^{24} \sin(\wJ T)
\sin^2(\tfrac{1}{4}\beta^{12})
+\ldots
\end{align}
which shows the excitation of double-quantum coherence as well as other orthogonal terms, in the general case (omitted for simplicity). 

This gives the following expression for the double-quantum excitation amplitude by the simplest implementation of SLIC given in Figure~\ref{fig:SLIC-DQ-sequences}(a):
\begin{align}
\label{eq:aDQSLICa}
a_\mathrm{DQ}^\mathrm{SLIC(a)}(T)
& = \bra{2}\rho_f(T)\ket{4}
\nonumber\\
&\simeq
i \sin(\wJ T)
\sin^2(\tfrac{1}{4}\beta^{12}).
\end{align}
where the SLIC rotation angle is given by Equation~\ref{eq:beta12SLIC}.

This compact analytical expression is compared with numerical calculations in Figure~\ref{fig:ExcitationTrajectories}(b). 
The SLIC double-quantum excitation amplitude is the product of two oscillating functions with different frequencies: A ``fast" oscillation with frequency $\wJ$, and a ``slow" oscillation with frequency $2^{-1/2}\wD$. Optimal double-quantum excitation requires setting $\tau$ to satisfy Equation~\ref{eq:tauSLICDQopt}, which maximises the value of the ``slow" oscillation, while simultaneously maximising the value of the ``fast" oscillation, by satisfying the condition:
\begin{equation}
\label{eq:tauSLICDQ(a)-cond}
\text{SLIC-DQ(a)}:\ 
T =  (k \pm \tfrac14)\, \tau_J
\ ;\quad k\in\mathbb{Z}.
\end{equation}
In general, the conditions of Equations~\ref{eq:tauSLICDQopt} and \ref{eq:tauSLICDQ(a)-cond} cannot simultaneously be satisfied exactly. A good compromise is to use a value of $T$ which satisfies Equation~\ref{eq:tauSLICDQ(a)-cond} exactly, while also satisfying Equation~\ref{eq:tauSLICDQopt} reasonably well. 

Note that double-quantum excitation is \emph{not} achieved when the duration $T$ is equal to a whole multiple of the period $\tau_J$; On the contrary, $T$ should be a whole multiple of $\tau_J$, \emph{plus or minus one extra quarter of a period} - a rather unusual constraint.  

The derivation above shows that double-quantum excitation by SLIC is a disguised form of Spinor-DQ excitation. Instead of the explicit $\pi/2$ pulses shown in Figure~\ref{fig:DQexc-schemes}(c) and 
\ref{fig:DQ-excitation-Spinor-DQ}, the $\pi/2$ rotations involved in SLIC are concealed in the interaction frame transformation. Nevertheless, the basic principles of Spinor-DQ excitation are evident. 

Equations~\mbox{\ref{eq:aDQSLICa}} and figure~\mbox{\ref{fig:ExcitationTrajectories}(b)} show that although a bare SLIC pulse does generate DQ coherence with high efficiency when applied to longitudinal magnetization, the dependence on the SLIC pulse duration is highly oscillatory.

Note that the sequence duration $T$ required to achieve SLIC-DQ excitation is twice that needed to achieve the generation of singlet order by SLIC\cite{devience_preparation_2013a}. 

\begin{figure}[tb]
\centering
\includegraphics[width=\columnwidth]{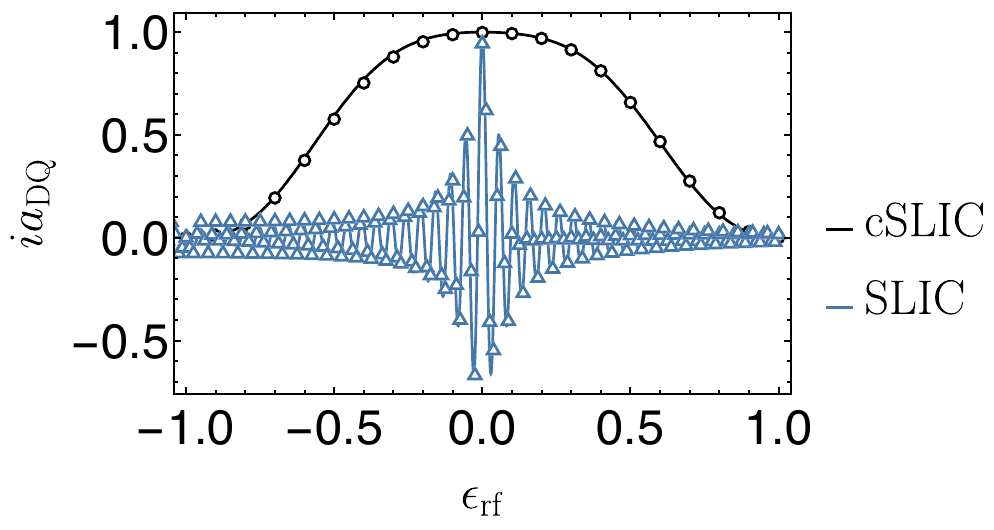}
\caption{
Double-quantum excitation as a function of rf amplitude error $\epsilon_{\rm{rf}}$ for the standard SLIC-DQ implementation shown in Figure~\ref{fig:SLIC-DQ-sequences}(a) as well as the cSLIC-DQ sequence shown in Figure~\ref{fig:SLIC-DQ-sequences}(c).
The total duration of the conventional SLIC-DQ sequence is $T = 79.12\mathrm{\,ms}$, corresponding to $T=(20+ \tfrac14)\tau_J$. For the compensated SLIC-DQ sequence, the total duration is $T = 78.14\mathrm{\,ms}$, corresponding to $T=20\tau_J$. The nutation frequency of the hard pulses in the cSLIC-DQ sequence was 25.0~kHz. The resonance offset is set to zero in both cases, $\omega_\Sigma=0$.  Solid lines: analytical expressions, as given by Equations~\ref{eq:aDQcSLIC} and~\ref{eq:aDQSLICwithepsilon}. 
Open markers: numerical simulations.
All simulations use $J = 255.94$~Hz and $\Delta=17.8$~Hz. 
}
\label{fig:SLIC-RFerrors-plot}
\end{figure} 
 
\subsubsection{SLIC-DQ: Two-pulse implementation}
The condition on the excitation duration $T$ given in Equation~\ref{eq:tauSLICDQ(a)-cond} is somewhat awkward. The version of SLIC-DQ shown in Figure~\ref{fig:SLIC-DQ-sequences}(b) corrects this minor problem. The SLIC pulse is followed by a strong $\pi/2$ pulse, \emph{with the same phase as the SLIC pulse}. In this case, the DQ excitation amplitude is given by:
\begin{equation}
\label{eq:aDQSLICb}
a_\mathrm{DQ}^\mathrm{SLIC(b)}(T)
\simeq
i \cos(\wJ T)
\sin^2(\tfrac{1}{4}\beta^{12}),
\end{equation}
where the SLIC rotation angle is given as usual by Equation~\ref{eq:beta12SLIC}. 

In this implementation, optimal double-quantum excitation is achieved when $T$ is an integer multiple of the period $\tau_J=|J^{-1}|$, while being as close as possible to the optimum value given in Equation~\ref{eq:tauSLICDQopt}. The SLIC duration is given by
\begin{equation}
\text{SLIC-DQ(b)}:\ 
    T = n_J\tau_J
    \ ;\ n_J \in \mathbb{Z},
\end{equation}
where the optimal number of cycles may be estimated as follows, in the absence of relaxation:
\begin{equation}
\label{eq:nJSLIC}
n_J 
\simeq
\mathrm{round}\!\left(
\left|
\frac{\sqrt{2}J}{\Delta}
\right|
\right).
\end{equation}

The double-quantum \emph{filtering} amplitude by SLIC-DQ is given by the square magnitude of Equation~\ref{eq:aDQSLICb}:
\begin{multline}
\label{eq:aDQFSLICb}
a_\mathrm{DQF}^\mathrm{SLIC(b)}(T)
=
|a_\mathrm{DQ}^\mathrm{SLIC(b)}(T)|^2
\\
\simeq
\cos^2(\wJ T)
\sin^4(\tfrac{1}{4}\beta^{12}).
\end{multline}

\emph{Rf amplitude dependence.}
Appendix~\ref{appx:SLICwithepsilon} gives a treatment of double-quantum excitation by SLIC, in the case that the rf amplitude deviates from the exact match condition ($\epsilon_\mathrm{rf}\neq 0$). Figure~\ref{fig:SLIC-RFerrors-plot} compares analytical results with numerical simulations of the double-quantum excitation, as a function of the rf field amplitude. A strong oscillatory dependence is superposed on the typically narrow excitation profile of SLIC. This renders SLIC-DQ extremely sensitive to small deviations of the rf amplitude from the exact SLIC match condition - even more sensitive than other applications of SLIC. As shown below, this hypersensitivity to the rf field amplitude strongly degrades the experimental performance of SLIC-DQ. 

\begin{figure}[tb]
\centering
\includegraphics[width=\columnwidth]{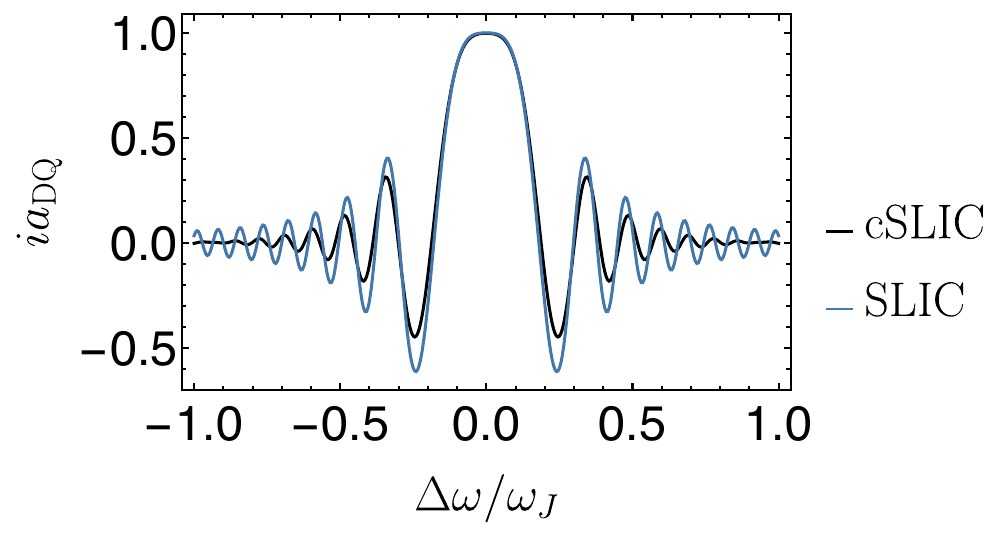}
\caption{
Numerical simulations of the performance of SLIC double-quantum excitation as a function of $\Delta\omega/\omega_J$, where the resonance offset  is defined as $\Delta\omega =\omega_\Sigma/2$. 
The total duration of the conventional SLIC-DQ sequence is $T = 79.12$ $\rm{ms}$, corresponding to $T=(20+\tfrac14)\tau_J$.
For the compensated SLIC-DQ sequence, the total duration was $T = 78.14$ $\rm{ms}$, corresponding to $T=20\tau_J$.
The nutation frequency of the hard pulses in the compensated SLIC-DQ sequence was 25.0~kHz, while the nutation frequency of the SLIC pulse was set to the nominal value $\omega_{\rm{nut}} = \omega_{J}$. 
All simulations use $J = 255.94$~Hz and $\Delta=17.8$~Hz. 
}
\label{fig:SLIC-detuning-plot}
\end{figure} 

\emph{Resonance offset dependence.}
Figure~\ref{fig:SLIC-detuning-plot} shows the performance of the SLIC-DQ sequence as a function of resonance offset $\Delta\omega=\tfrac12\omega_\Sigma$. Exact SLIC match is assumed, $\epsilon_\mathrm{rf}=0$.  The simulation shows that double-quantum excitation by SLIC is strongly frequency-selective, with an excitation bandwidth roughly equal to $0.3$ times the $J$-coupling. 

\subsubsection{cSLIC-DQ: Amplitude-Error Compensation}
\label{sec:cSLIC}

The high sensitivity of SLIC to deviations in the rf field amplitude from the exact SLIC condition is a serious impediment to its practical use, especially in the context of double-quantum excitation. Fortunately, an effective scheme for rf error compensation is available. 

In many NMR experiments, pulse sequences are compensated for rf amplitude errors by introducing a $\pi$ phase shift of the rf field, so that a net positive nutation under the resonant field is compensated by an equal net negative rotation. For example, cross-polarization sequences may be compensated this way~\cite{levitt_spin_1986,levitt_heteronuclear_1991}. However, the same method does not work in the context of SLIC, since a $\pi$ phase shift in the SLIC irradiation field switches the single-transition rotation from the $\{\ket{1},\ket{2}\}$ subspace to the $\{\ket{1},\ket{4}\}$ subspace. The operation of the SLIC sequence is therefore completely disrupted by a $\pi$ phase shift of the SLIC-matched rf field. 

An alternative method, which compensates SLIC for rf amplitude variations without disruption, is shown in Figure~\ref{fig:SLIC-DQ-sequences}(c). This method, which is called here compensated-SLIC (cSLIC), uses an rf pulse sequence employing two different amplitude levels: a weak rf field matching the SLIC condition, and a much stronger rf field which provides a compensating counter-rotation in the centre of each repeating element. 

The general form of the cSLIC sequence, taking into the finite duration of the central strong pulse, has the form:
\begin{equation}
\label{eq:cSLIC}
\text{cSLIC} = (\alpha \pi)_x - (\alpha 2\pi)_{-x}^\mathrm{strong}-(\alpha\pi)_{x},
\end{equation}
where the factor $\alpha$ is close to $1$ and is given by:
\begin{equation}
\label{eq:alpha factor}
\alpha = \frac{\omega^\mathrm{strong}_\mathrm{nut}}{
\omega^\mathrm{strong}_\mathrm{nut}+\omega_J
}.
\end{equation}
The total duration of the cSLIC element is $\tau_J=|J|^{-1}$. Here $\omega^\mathrm{strong}_\mathrm{nut}$ is the nutation frequency of the strong pulse, $\omega^\mathrm{strong}_\mathrm{nut}\gg\omega_J$. The nutation frequency of the two weak outer pulses matches the SLIC condition, $\omega_\mathrm{nut}=\omega_J$.

Consider the limit where the duration of the central compensating pulse is negligible compared to that of the SLIC pulse elements. In this limit, $\alpha=1$, and
each of the $n_J$ cycles shown in Figure~\ref{fig:SLIC-DQ-sequences}(b) is divided into two equal intervals, each of duration $\tau_J/2$. In this limit, the two weak pulses each have a flip angle of $\pi$, while the strong and short central pulse has a flip angle of $2\pi$. The phase of the strong $2\pi$ pulse is shifted by $\pi$, with respect to the phase of the weak SLIC-matched field. 

The sequence of two SLIC-matched $\pi$ pulses, bracketing one strong $2\pi$ pulse with a phase shift of $\pi$, has a total duration of $\tau_J=|J^{-1}|$. This cycle is repeated $n_J$ times, where $n_J$ may be estimated through Equation~\ref{eq:nJSLIC}, ignoring relaxation. 

The cSLIC sequence is compensated for any excess nutation caused by a misset rf field amplitude. Under ideal conditions, the weak SLIC field generates an exact $2\pi$ rotation upon every cycle, and the central pulse generates an equal and opposite $2\pi$ rotation. When the rf field amplitude is larger than expected, the excess rotation induced by the misset SLIC field is  compensated by the equal and opposite excess rotation caused by the misset strong pulse -- and similarly when the rf field is weaker than nominal. This compensation mechanism assumes that the weak and strong rf fields experience the rf amplitude errors in the same proportion. This is the case when both fields are generated by the same radiofrequency coil, and when the deviations are caused by spatial variations in the rf field strength. This is the usual experimental situation. 

As discussed in Appendix~\ref{appx:cSLICtheory}, the cSLIC scheme of Figure~\ref{fig:SLIC-DQ-sequences}(c) removes the second term (the rf error term) from the average Hamiltonian of Equation~\ref{eq:HbarSLIC}. The dependence of the cSLIC-DQ excitation amplitude on the total sequence duration $T$ and rf amplitude error $\epsilon$ is given by:
\begin{equation}
\label{eq:aDQcSLIC}
a_{\mathrm{DQ}}^\mathrm{cSLIC}
\simeq 
+i \cos\theta_{\epsilon}
\sin^2\!\left(
\tfrac{1}{4}\kappa'\omega_{\Delta} T\right) \ ,
\end{equation}
where the parameters $\theta_{\epsilon}$ and $\kappa'$ are defined as follows:
\begin{align}
\label{eq:thetaepsilon+kappap}
\theta_{\epsilon} &= 2\,\arctan\left(
\frac{\epsilon_{\mathrm{rf}}}{2+\epsilon_{\mathrm{rf}}}
\right), \nonumber \\
\kappa'&= 
2^{-1/2}
\mathrm{sinc}\left(\pi \epsilon_{\mathrm{rf}}\right)
\sec\left(\theta_{\epsilon}\right). 
\end{align}
In the case of exact SLIC match ($\epsilon_{\mathrm{rf}}\to0$), these parameters tend to the values $\theta_\epsilon\to0$ and $\kappa'\to\kappa$, so that cSLIC has the same behaviour as SLIC. 

This leads to the double-quantum-filtered signal amplitude:
\begin{multline}
\label{eq:aDQFcSLIC}
a_\mathrm{DQF}^\mathrm{cSLIC}(T)
=
|a_\mathrm{DQ}^\mathrm{cSLIC}(T)|^2
\simeq
\cos^2(\theta_\epsilon)
\sin^4(\tfrac{1}{4}\kappa'\omega_{\Delta}T).
\\
\end{multline}

The strong-pulse limit with $\alpha=1$ can only be approached if the duration of the central pulse is negligible, which requires a very strong radiofrequency field amplitude. A compromise for the finite-pulse case, which works well in practice, is to slightly reduce the durations, and hence the flip angles, of the two outer pulses, while simultaneously reducing the flip angle of the central strong pulse, in order to match the sum of the flip angles of the outer pulses. The overall duration of the three pulses remains fixed at $\tau_J$.  This leads to the compromise solution in equation~\ref{eq:cSLIC} and figure~\ref{fig:SLIC-DQ-sequences}(c).

The greatly improved robustness of cSLIC-DQ with respect to rf amplitude errors is shown by the  numerical simulations  in Figure~\ref{fig:SLIC-RFerrors-plot}. The resonance-offset dependence of uncompensated SLIC-DQ and cSLIC-DQ are similar, as shown by the simulations in Figure~\ref{fig:SLIC-detuning-plot}.

\section{Materials and Methods}
\label{sec:Methods}

\subsection{Calculations}
All numerical simulations, and analytical results, were generated by \emph{SpinDynamica} (\texttt{www.spindynamica.soton.ac.uk})~\cite{bengs_spindynamica_2018}.  

\begin{table}[tb]
\begin{center}
\caption{
The molecular system used for the experiments, tert-butyl (2,2-difluoro-2-phenylacetyl)-L-alaninate, referred to as \textbf{I}, and associated parameters. 
The values of $\Delta$ and $\thst$ assume a field of 14.1~T.
}
\renewcommand{\arraystretch}{1.5}
\begin{tabular}{c c c} \hline
\multicolumn{2}{c}{\adjustbox{valign=m,margin=0pt 3pt 0pt 3pt}{\includegraphics[width=0.65\linewidth]{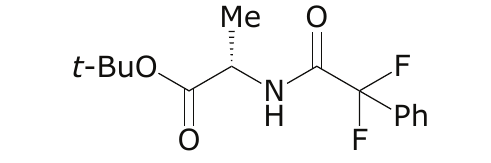}}} \\ \hline
 \textbf{Parameter} & \textbf{Value} \\ \hline
 $^{2}J_{\text{FF}}$ /Hz & $255.9 \pm 0.3$ \\ 
$\Delta\delta$ /ppb & $32 \pm 2$ \\ \hline
 $\Delta$ /Hz & 18 $\pm 1$ \\
  $\thst /\degree$ & $4.1\pm0.1$ \\ 
\hhline{= =}

\label{table: fluorinated amide structure and parameters}
\end{tabular}
\end{center}
\end{table}

\subsection{Sample Preparation}
\label{sec:sample}

$^{19}\mathrm{F}$ NMR experiments were performed on the compound tert-butyl (2,2-difluoro-2-phenylacetyl)-L-alaninate, referred to here as \textbf{I}. The molecular structure of \textbf{I}, and the relevant NMR parameters, are shown in Table~\ref{table: fluorinated amide structure and parameters}. This compound contains a diastereotopic pair of \F{19} nuclei. The spin-system parameters were estimated from the NMR response to $J$-synchronized pulse sequences \cite{pileio_longlived_2012,sheberstov_excitation_2019}.

The synthetic procedure for \textbf{I} is given in the Supplementary Material. 

The experiments used $\uL{400}$ of a 
$\sim$0.15 M solution of \textbf{I} in CDCl\textsubscript{3}, loaded into a susceptibility-matched Shigemi tube. 

\begin{table}[b]
\begin{center}
\caption{Pulse sequence parameters used for the double-quantum filtering experiments on the solution of \textbf{I}. 
The GeoDQ parameters $\{\tau_{1}^\mathrm{Geo}, \tau_{2}^\mathrm{Geo},n\}$ correspond to Figure~\ref{fig: geodq pulse sequence}. 
The factor $\alpha$ for cSLIC is defined in Equation~\ref{eq:alpha factor}. The uncertainties in the nutation frequencies were estimated from the experimental nutation spectrum.
}
\renewcommand{\arraystretch}{1.5}
\begin{tabular}{c c c}
 \textbf{Pulse Sequence} & \textbf{Parameter} & \textbf{Value} \\ \hline
Hard pulses & $\omega_\mathrm{nut}/(2\pi)$ / kHz & $25.0 \pm 0.3$ 
\\ \hline
INADEQUATE & $\tau_1$ / ms & $145.00$ 
\\& $T$ / ms & 290.00
\\ \hline
\multirow{3}{*}{GeoDQ} & $\tau_{1}^\mathrm{Geo}$ / $\mu$s & $926.0$ \\ 
& $\tau_{2}^\mathrm{Geo}$ / $\mu$s & $1852$ \\
& $n$ & $22$ \\
& $T$ / ms& $85.88$
\\ \hline 
\multirow{2}{*}{PulsePol-DQ} 
& $\tau_{2}$ / $\mu$s 
& $1407.5$ \\
& $m$ & $9$
\\& $T$ / ms & $105.66$
\\ \hline 
SLIC-DQ & $\omega_\mathrm{nut}^\mathrm{SLIC}/(2\pi)$ / Hz & $256  \pm 3$ \\
& $T$ / ms & $70.866$
\\ \hline
\multirow{4}{*}{\shortstack{cSLIC-DQ}}
& {$\omega_\mathrm{nut}^\mathrm{SLIC}/(2\pi)$} / Hz & $256 \pm 3$ \\
& $n_J$ & $19$ \\ 
& $\tau_J$ / $\mu$s & $3916$ \\
& $T$ / ms& $74.404$ 
\\
& $\alpha$ & $0.990$
\\
\hhline{= = =}
\end{tabular}
\label{table: PulseSequenceParameters}
\end{center}
\end{table}

\subsection{NMR Instrumentation}
\label{sec:Instrumental}

\textsuperscript{19}F spectra were acquired at a magnetic field strength of 14.1 T using a Bruker Avance NEO spectrometer and a custom Bruker 5 mm TBO iProbe with four channels ($^{1}$H/$^{2}$H/$^{3}$He/BBF). 

Unless otherwise specified, the pulse powers were calibrated to provide a $^{19}\mathrm{F}$ nutation frequency of $\omega_{\mathrm{nut}} = 2\pi \times 25 \mathrm{kHz}$ corresponding to a $90\degree$ 
pulse length $\tau_{90}$ of $10~\mu\text{s}$. 

No $\mathrm{^1H}$ decoupling was used throughout the experiments. 
 
The \F{19} $T_1$ relaxation time of the \textbf{I} solution was estimated by inversion recovery to be $746\pm 3$ ms at $14.1$~T field strength, as shown in the Supporting Information. 

All double-quantum-filtering pulse sequences consisted of
a double-quantum excitation block with overall phase $\Phi_A$, followed by a double-quantum reconversion block with overall phase $\Phi_B$. Quadrature detection of the signal was conducted with a receiver/digitizer phase $\Phi_\mathrm{rec}$. Double-quantum coherence at the junction of the excitation and reconversion blocks was selected by cycling the phases of the blocks in a standard four-step cycle: $\Phi_{\text{A}} = \{0,0,0,0\}$, $\Phi_{\text{B}} =\{0,\pi/2,\pi,3\pi/2\}$ and $\Phi_{\text{rec}} = \{0,3\pi/2,\pi,\pi/2\}$. For the PulsePol-DQ scheme, an additional phase shift of $\pi$ was applied to $\Phi_{\text{B}}$. All spectra were processed identically. 

In all double-quantum filtered experiments, an additional magnetic field gradient pulse was inserted, in the form of a $z$-filter, to suppress residual coherences before the final readout pulse. The gradient pulse strength was $20$ G cm$^{-1}$. 

An empirically calibrated phase adjustment of $-1.4 \degree$ was applied when switching between the strong and weak pulses of the cSLIC sequence. 

\begin{figure}[t]
\centering
\includegraphics[width=0.75\columnwidth]{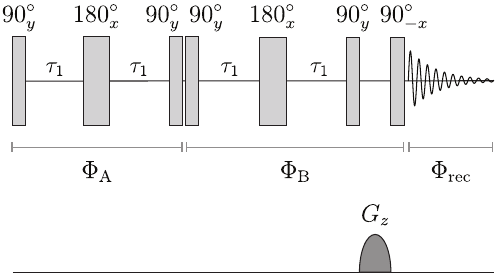}
\caption{Pulse sequence for double-quantum filtering using the refocussed INADEQUATE scheme, with a $z$-filtering step before acquisition. The phases $\Phi_\mathrm{A}$, $\Phi_\mathrm{B}$ and $\Phi_\mathrm{rec}$ are cycled in 4 steps to implement double-quantum filtering.}
\label{fig: refocused inadequate pulse sequence}
\end{figure} 

\begin{figure*}[hbt!]
\centering
\includegraphics[width=0.75\linewidth]{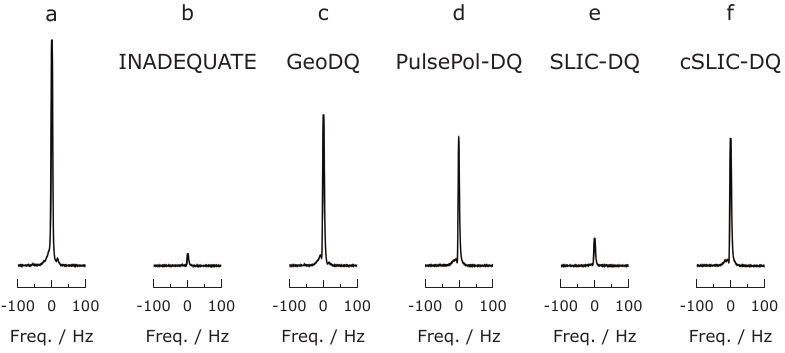}
\caption{
\F{19} spectra of a solution of \textbf{I} in CDCl\textsubscript{3}, acquired at $14.1$ T, $298$ K, and averaged over 4 transients. $\mathrm{^1H}$ decoupling was not used. The frequency axis is centred around $-102.931$ ppm. All spectra have the same vertical scale. All pulse sequence parameters are given in Table~\ref{table: PulseSequenceParameters}. Double-quantum-filtering amplitudes are given below in parentheses. 
({a}) A $90\degree$ pulse-acquire spectrum.
({b}) Double-quantum-filtered spectrum obtained using the refocused INADEQUATE pulse sequence (Figure~\ref{fig: refocused inadequate pulse sequence}) [5.4\%]. 
({c}) Double-quantum-filtered spectrum obtained using the GeoDQ pulse sequence (Figure~\ref{fig: geodq pulse sequence})  [66.6\%].  
({d}) Double-quantum-filtered spectrum obtained using the PulsePol-DQ pulse sequence (Figure~\ref{fig: pulsepol dq pulse sequence}) [56.5\%].  
({e}) Double-quantum-filtered spectrum obtained using the SLIC-DQ pulse sequence (Figure~\ref{fig: slic dq Gz pulse sequence}) [15.0\%].  
({f}) Double-quantum-filtered spectrum obtained using the cSLIC-DQ pulse sequence (Figure~\ref{fig: slic dq Gz pulse sequence}) [56.2\%].
}
\label{fig:ExperimentalDQFSpectra}
\end{figure*}

\section{Results}
\label{sec:Results}
A set of experimental \textsuperscript{19}F NMR spectra of the solution of \textbf{I} 
is shown in Figure~\ref{fig:ExperimentalDQFSpectra}. 
The corresponding pulse sequence parameters are given in Table~\ref{table: PulseSequenceParameters}.

Figure~\ref{fig:ExperimentalDQFSpectra}({a}) shows the ordinary pulse-acquire spectrum obtained by a Fourier transform of the free-induction decay generated by a $90\degree$ excitation pulse. This spectrum shows only a single strong peak, without any obvious sign of magnetic inequivalence for the \F{19} pair.

Figures~\mbox{\ref{fig:ExperimentalDQFSpectra}({b-f})} show double-quantum-filtered \mbox{\F{19}} spectra of \textbf{I} for a variety of double-quantum excitation schemes. Control experiments were also performed using the same pulse sequences on a sample with a magnetically equivalent \mbox{\F{19}} pair. As expected, the double-quantum-filtered signals are of negligible intensity in this case (see Supporting Information). This proves that the signals shown in figures~\mbox{\ref{fig:ExperimentalDQFSpectra}({b-f})} did indeed pass through a double-quantum state.

\begin{figure}[b]
\centering
 \includegraphics[width=0.75\columnwidth]{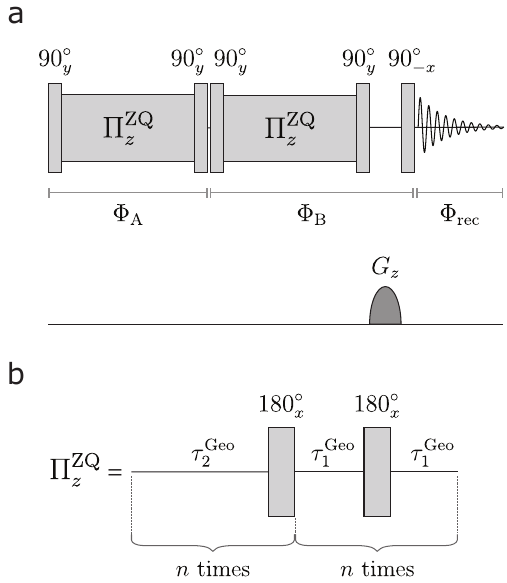}
\caption{(a) Double-quantum filtering pulse sequence, using GeoDQ for double-quantum excitation, with a $z$-filtering step before acquisition. 
(b) Structure of the $\Pi_z^\mathrm{ZQ}$ block.
Theoretical expressions for $\tau_1^{\rm{Geo}}$, $\tau_2^{\rm{Geo}}$ and $n$ are given in Ref.~\cite{bengs_aharonov_2023}. The phases $\Phi_\mathrm{A}$, $\Phi_\mathrm{B}$ and $\Phi_\mathrm{rec}$ are cycled in 4 steps to implement double-quantum filtering.}
\label{fig: geodq pulse sequence}
\end{figure} 

\subsubsection{Refocused INADEQUATE}
Figure~\ref{fig:ExperimentalDQFSpectra}({b}) shows a double-quantum-filtered \F{19} NMR spectrum obtained using the refocused INADEQUATE pulse sequence, as illustrated in Figure~\ref{fig: refocused inadequate pulse sequence}. Even after optimisation of the pulse sequence delays $\tau_1$, only $\approx 5\%$ of the magnetization is retained after passing through the double-quantum filter.

Figure~\ref{fig:DQtrajectories}(a) shows 
the experimental dependence of the double-quantum filtered signal intensity on the sequence duration $T$ for the INADEQUATE pulse sequence. Whereas the theoretical curve (blue line) and simulated points (open symbols) continue to increase at large $T$ values, the experimental amplitudes (black line) are heavily damped, shortening the optimal $\tau_1$, and greatly reducing the optimal double-quantum signal. As shown in table~\ref{tab:DQExcitationDurations}, the experimental optimum duration for INADEQUATE is much shorter than the theoretical optimum in the absence of relaxation. 

\subsubsection{GeoDQ}
Figure~\ref{fig:ExperimentalDQFSpectra}({c}) shows the double-quantum filtered spectrum obtained \textit{via} the GeoDQ pulse sequence, given in Figure~\ref{fig: geodq pulse sequence}, with the parameters given in table~\ref{table: PulseSequenceParameters}. In this case, approximately $60\%$ of the magnetization is retained after passing through the double-quantum filter. This is roughly $12$ times more than that generated by INADEQUATE. The improved efficiency may be attributed to the much faster double-quantum excitation by GeoDQ, compared to that achieved by INADEQUATE.

\begin{table}[tb]
\begin{center}
\caption{Total pulse-sequence durations $T$ for double-quantum excitation in strongly-coupled 2-spin-1/2 systems. The theoretical values are given in the near-equivalence limit $\vert{\Delta}\vert \ll \vert{J}\vert$, using the spin system parameters for \textbf{I}, and neglecting relaxation. The last column shows the experimentally optimised values for the solution of \textbf{I}. 
}
\renewcommand{\arraystretch}{1.5}
\begin{tabular}{c c c}
\textbf{Sequence} & \textbf{$T$ (theory)} & \textbf{$T$ (exp)} \\ \hline
INADEQUATE & $J/\Delta^2 = 808.00\,\text{ms}$ & $290.00\,\text{ms}$ \\[15pt] 
GeoDQ & $1.571/\Delta = 88.25\,\text{ms}$ & $85.88\,\text{ms}$\\[15pt]
PulsePol-DQ & $1.952/\Delta = 109.66\,\text{ms}$ & $105.55\,\text{ms}$ \\[15pt]
SLIC-DQ & \multirow{2}{4em}{$1.414/\Delta = 79.44\,\text{ms}$} & $70.9\,\text{ms}$ \\
cSLIC-DQ & & $74.4\,\text{ms}$ \\
\hhline{= = =}
\end{tabular}
\label{tab:DQExcitationDurations}
\end{center}
\end{table}

\subsubsection{PulsePol-DQ}
Figure~\ref{fig:ExperimentalDQFSpectra}({d}) shows the double-quantum filtered spectrum obtained using the PulsePol-DQ pulse sequence, as shown in Figure~\ref{fig: pulsepol dq pulse sequence}. The double-quantum-filtered signal amplitude is again much greater than for INADEQUATE, but is slightly less than that achieved by GeoDQ. This is attributed to PulsePol-DQ having a longer excitation duration $T$ than GeoDQ (see Table~\ref{tab:DQExcitationDurations}). 

Figure~\ref{fig:DQtrajectories}(b) compares 
the experimental $T$-dependence of the double-quantum-filtered signal intensity for the PulsePol-DQ pulse sequence (black curve) with the analytical formula (blue line) and simulated values (open symbols). There is a good correspondence between theory and experiment, except for the obvious damping of the curves by $T_2$ relaxation. The experimentally optimized value of $T$ is shown by a dashed line. This corresponds closely to the theoretical optimum (see Table~\ref{tab:DQExcitationDurations}).

\subsubsection{SLIC-DQ}
The pulse sequence shown in Figure~\ref{fig: slic dq Gz pulse sequence} implements double-quantum-filtered NMR using SLIC. 
Figure~\ref{fig:ExperimentalDQFSpectra}({e}) shown the double-quantum filtered spectrum obtained, when the uncompensated SLIC sequence in Figure~\ref{fig:SLIC-DQ-sequences}(b) is used. The double-quantum-filtered signal is clearly present, but its amplitude is small, even after experimental optimization. 

Figure~\ref{fig:DQtrajectories}(c) shows 
the experimental dependence of the double-quantum filtered signal intensity on the sequence duration $T$ for the SLIC-DQ pulse sequence. Although SLIC-DQ is faster than GeoDQ and PulsePol-DQ, the final DQF spectrum only retains approximately $15\%$ of the magnetization, in the best case. The strong losses are attributed to the extreme sensitivity of uncompensated SLIC-DQ to the rf amplitude deviations, as illustrated by the simulations in  Figure~\ref{fig:SLIC-RFerrors-plot}. Small variations of the rf field amplitude across the sample lead to premature decay of the double-quantum-filtered signal. 

\subsubsection{cSLIC-DQ}
The large signal loss of SLIC-DQ due to rf inhomogeneity is partially recovered by the compensated cSLIC-DQ sequence of Figure~\ref{fig:SLIC-DQ-sequences}(c). 
The optimised double-quantum-filtered cSLIC-DQ spectrum is shown in Figure~\ref{fig:ExperimentalDQFSpectra}({f}) and has a comparable intensity to that achieved by PulsePol-DQ. The double-quantum excitation time $T$ for cSLIC-DQ is the shortest of all the sequences explored here (see table~\ref{tab:DQExcitationDurations}). 

The experimental $T$-dependence of the double-quantum-filtered signal for cSLIC-DQ is shown by the black solid line in Figure~\ref{fig:DQtrajectories}(d). The initial part of this curve corresponds well to the analytical theory (blue curve) and the simulated points (open symbols), when the obvious damping of the curves due to relaxation is taken into account. However, at longer times $T$, the experimental curve deviates appreciably from theory, even going negative for $T\sim 0.25\mathrm{\,s}$.

The reasons for this deviation are currently unknown. Instrumental droop in the amplitude of the strong compensation pulses over a long pulse train may be responsible. The operation of the sequence depends critically on the exact balance of the strong compensation pulses and the weak SLIC-matched field. If the power of the compensation pulses declines as the sequence proceeds, with the weak SLIC-pulse amplitudes remaining constant, a small net rotation would accumulate on each cSLIC repetition. This could explain the observed behaviour. 
Another possibility is the accumulating effect of phase or amplitude transients at the leading or falling edges of the pulses. Such effects are well-documented, especially in the context of solid-state NMR~\cite{mehring_phase_1972,haeberlen_line_1977,prigl_theoretical_1996,barbara_phase_1991}. Compensation strategies have been devised~\cite{vega_controlling_2004,skinner_reducing_2004,takeda_active_2009,tabuchi_total_2010,weber_pulsetransient_2012,wittmann_compensating_2015,goodwin_advanced_2017,stasiuk_frame_2024}. 

\section{Discussion}

Double-quantum excitation and double-quantum filtering are widely-used techniques in solution NMR, allowing the simplification of NMR spectra, the suppression of undesired signals from isolated nuclei~\cite{bax_natural_1980a,bax_nmr_1981}, and the correction of phase errors in multidimensional spectra~\cite{rance_DQFCOSY_1983}. Although most existing double-quantum excitation methods are well-adapted for the case of widely separated chemical shifts (the ``weak coupling") case, the work in this paper and our previous one~\cite{bengs_aharonov_2023} has shown that techniques may also be developed for the case of near-equivalence or extreme strong coupling. Both sets of methods explore special features of quantum dynamics. The method of 
geometric double-quantum excitation, described in Ref.~\cite{bengs_aharonov_2023} exploits the Aharanov-Anandan geometric phase. This is a phase factor acquired by quantum states when they are taken around a cyclic trajectory that subtends a finite solid angle at the origin of the Bloch sphere. 
The current paper, on the other hand, exploits spinor behaviour, which implies that a quantum state changes sign when rotated by $2\pi$ in a two-dimensional subspace.  As described above, either one of these phenomena may be leveraged to manipulate strongly coupled spin-1/2 pairs, such that double-quantum coherence is excited in a time much shorter than is possible by standard methods. 

Two main varieties of spinor double-quantum excitation have been described. In the first method, repetitions of the PulsePol sequence,
which is interpreted here using symmetry-based recoupling theory~\cite{sabba_symmetrybased_2022}, 
is used to generate a single-transition rotation through an angle of $2\pi$, which realizes the spinor property and allows the efficient excitation of double-quantum coherence. In the second method, a single radiofrequency pulse matched to the SLIC condition~\cite{devience_preparation_2013a} is used. As shown above, this method may be interpreted as a disguised form of Spinor-DQ excitation. 

As far as we know, the direct excitation of double-quantum coherence by SLIC has not been described before, although related multiple-quantum excitation effects have been noted when double-frequency radiofrequency fields are applied to coupled spin systems~\cite{emsley_double_1990,elbayed_behaviour_1990,zhou_response_1996,vincent_selective_1993,konrat_coherence_1991,konrat_specific_1994}. 

Some variants of SLIC-DQ excitation were introduced. These include a repeating three-pulse sequence, using two different radiofrequency field amplitudes, which we call cSLIC. This is a variety of SLIC which is well-compensated for deviations of the rf field amplitude from the exact SLIC condition, although effects such as radiofrequency droop and phase transients remain problematic after multiple repetitions.

These methods were demonstrated by the \F{19} NMR of a molecular system containing a diastereotopic $\mathrm{CF_2}$ moiety, where a small chemical shift difference between the \F{19} nuclei is induced by a stereogenic centre elsewhere in the same molecule. In this case, the chemical shift difference is too small to induce a resolved splitting in the \F{19} NMR spectrum, which has the appearance of a single peak. Nevertheless, all of the methods explored here provided good double-quantum excitation, even in this unfavourable case, with two exceptions. The first exception was the conventional three-pulse INADEQUATE method, which performed poorly, as expected for this extreme strong-coupling case. The second exception was uncompensated SLIC-DQ, 
which also performed poorly, presumably because of its extreme sensitivity to radiofrequency field inhomogeneity; the SLIC-DQ method performed as well as the others when this defect was corrected by using the cSLIC-DQ scheme. 

The geometric double-quantum excitation method has already been used for the selective detection of \Ctwo signals while suppressing much larger signals from isolated \Cth nuclei~\cite{heramun_singlet_2025}. Further applications of this kind may be anticipated for the GeoDQ and also the Spinor-DQ methods. For example, diastereotopic $\mathrm{^1H}$ pairs are common in the NMR of biomolecules such as proteins. The glycine residues in flexible peptide backbones, such as found in flexible termini or loops, contain inherently diastereotopic $\mathrm{CH_2}$ groups. In some cases, the chemical shift difference between the diastereotopic $\mathrm{^1H}$ nuclei are too small to give  a resolved spectral splitting. Nevertheless, the methods described here are capable of generating double-quantum coherence even in this regime, as is attested by the results shown in Figure~\ref{fig:ExperimentalDQFSpectra}. 

The performance of GeoDQ is similar to that of PulsePol-DQ for the molecular system studied here, with GeoDQ having a slight edge in signal amplitude. PulsePol-DQ is somewhat more straightforward to implement, since the sequence is repetitive, which permits ready optimization of the overall sequence duration, as shown by the experimental results in fig.~\ref{fig:ExcitationTrajectories}(b). 

The cSLIC-DQ sequence is another viable alternative. cSLIC-DQ provides faster double-quantum excitation than the other sequences, which is advantageous in situations with rapid transverse relaxation. Its strong dependence on resonance offset, shown in figure~\ref{fig:SLIC-detuning-plot}, may be an advantage or a disadvantage, depending on the context. One point of concern is that cSLIC-DQ is susceptible to accumulating pulse errors over many repetitions, as shown by the experimental results in figure~\ref{fig:DQtrajectories}(d). 

Double-quantum-filtering using the methods described above could allow the editing of NMR spectra, as an alternative to techniques that exploit long-lived singlet states~\cite{mamone_gcM2S_2020,devience_nuclear_2013}. \emph{In vivo} and MRI applications are conceivable~\cite{mamone_localized_2021,lysak_vivo_2023,boele_ultralow_2025,mcbride_scalable_2025}.

The sequences described here are also likely to be applicable to systems of more than 2 coupled spins. For example, various implementations of SLIC have been applied successfully for the study of long-lived states in systems of the kind $\mathrm{AA'XX'}$ and $\mathrm{AA'MM'XX'}$~\cite{sonnefeld_polychromatic_2022,sonnefeld_longlived_2022,sheberstov_collective_2024,razanahoera_hyperpolarization_2024,wiame_longlived_2025}. 

The cSLIC sequence retains the frequency-selectivity of SLIC, while displaying high robustness with respect to rf amplitude deviations. The cSLIC sequence might be useful in other experiments which use SLIC, such as the manipulation of long-lived nuclear spin order~\cite{devience_preparation_2013a,
devience_nuclear_2013,
elliott_longlived_2016,
elliott_longlived_2016a,
eills_singlet_2017,
sheberstov_cis_2018,
elliott_nuclear_2019,
sheberstov_excitation_2019,
sheberstov_generating_2019,
devience_manipulating_2020,
korenchan_31p_2022,
sonnefeld_polychromatic_2022,
sonnefeld_longlived_2022,
razanahoera_hyperpolarization_2024,
sheberstov_collective_2024,
wiame_longlived_2025,
mandzhieva_zerofield_2025,
mcbride_scalable_2025}, and in parahydrogen-enhanced NMR~\cite{pravdivtsev_chemical_2018,knecht_efficient_2019,boele_ultralow_2025,mcbride_scalable_2025}. 

Another potential application of these techniques is for molecular binding assays. It has been found that, in some cases, the reversible association of an achiral molecule with a chiral one induces diastereotopicity in 
moieties of 
the achiral partner~\cite{ishihara_nmr_2018}. In principle, this phenomenon could be detected by double-quantum-filtered NMR experiments on the free achiral molecule in solution. As demonstrated above, double-quantum filtering is discriminatory even when the ordinary NMR spectrum displays no obvious symptom of diastereotopicity. Double-quantum filtering, using GeoDQ or Spinor-DQ methods, could therefore detect the reversible association of some types of molecules in solution. In the context of drug discovery, this could provide a complement to existing NMR-based methods, which typically exploit either transient magnetization transfer in the bound complex, enhancements in the decay rates of magnetization or long-lived spin order, or binding-induced perturbations of diffusion rates and chemical shifts~\cite{dalvit_identification_2000,stockman_nmr_2002,dalvit_ligand_2007,buratto_drug_2014,buratto_exploring_2014,buratto_ligand_2016}. Double-quantum filtering might provide a cleaner and less ambiguous discrimination in some cases. This possibility is under investigation.  

\section*{Appendices}
\appendix
\renewcommand{\thesection}{\Alph{section}}
\counterwithin{equation}{section}

\section{Higher order terms of the SLIC effective Hamiltonian}
\label{appx:SlicHigherOrderAvHam}

In general, the SLIC average Hamiltonian $\overline{H}_\mathrm{SLIC}^{(k)}$ ($k \in \mathbb{Z}$) contains higher order terms, which play a role in spin-1/2 pairs beyond the strong-coupling limit. The first few terms may be written as a power series of the variable $\zeta = \Delta/(\sqrt{2}J)$:

\begin{align}{\label{eq:HigherOrderAvHams}}
\overline{H}_\mathrm{SLIC}^{(2)} &= \frac{\omega_{J}\zeta^2}{4}\left(I_x^{24}-I_z^{14}\right) \nonumber \\
\overline{H}_\mathrm{SLIC}^{(3)} &= \frac{\omega_{J}\zeta^3}{16}\left(I_x^{12}+I_x^{14}\right) \nonumber \\
\overline{H}_\mathrm{SLIC}^{(4)} &= \frac{\omega_{J}\zeta^4}{256}\left(4I_z^{12}-I_{z}^{14}-8I_x^{24}+\pi I_y^{24}\right)
\end{align}

The effective Hamiltonian terms have the following features:
\begin{itemize}
    \item The odd-order average Hamiltonian terms  $\overline{H}_\mathrm{SLIC}^{(2k+1)}$ (with $k\in\mathbb{Z}$) contain singlet-outer-triplet shift operators within the subspaces $\{\ket{1},\ket{2}\}$ and  $\{\ket{1},\ket{4}\}$.
    \item The even-order average Hamiltonian terms $\overline{H}_\mathrm{SLIC}^{(2k)}$ contain longitudinal singlet-outer triplet detuning terms $\left\{I_z^{12},I_{z}^{14}\right\}$, playing a role similar to that of the Bloch-Siegert shift~\cite{abragam_principles_1983}, as well as 
    double-quantum terms $\left\{I_x^{24},I_{y}^{24}\right\}$. 
\end{itemize}

\section{SLIC double-quantum excitation in the presence of rf amplitude errors}
\label{appx:SLICwithepsilon}

Earlier in the paper, the effective Hamiltonian of the conventional (or "uncompensated") SLIC sequence $\overline{H}^{(1)}$ - including the rf amplitude error term $\epsilon_{\rm{rf}}$ - was given in Equation \ref{eq:HbarSLIC}, but we only considered the case $\epsilon_{\rm{rf}} = 0$. We now examine  the effect of  $\epsilon_{\rm{rf}}$ on double-quantum excitation. 

In the presence of $\epsilon_{\rm{rf}}$, the SLIC Hamiltonian of Equation \ref{eq:HbarSLIC} can be conveniently parametrized in the following form: 
\begin{equation}
\overline{H}^{(1)}
=
\Omega_{12} \left(\cos\left(\gamma_{\epsilon}\right) I_x^{12} + \sin\left(\gamma_{\epsilon}\right)I_z \right)
\end{equation}
where the detuning angle $\gamma_{\epsilon}$ and the effective Rabi frequency $\Omega_{12}$ are defined as follows:
\begin{align}
\gamma_{\epsilon} &= \arctan\left(\sqrt{2}\epsilon_{\rm{rf}} J/\Delta\right) \nonumber \\
\Omega_{12} &= 2^{-1/2}\omega_{\Delta}\sec
\left(\gamma_{\epsilon}\right)
\end{align}

Examining the form of the detuning angle $\gamma_{\epsilon}$ gives some insight to the root cause of the poor performance of the SLIC sequence: the term $\epsilon_{\rm{rf}}$ is multiplied by the ratio $J/\Delta$, such that strongly-coupled spin systems become \emph{more sensitive} to rf amplitude errors.

The Hamiltonian $\overline{H}^{(1)}$ may be inserted into $U(T)$, the propagator of Equation \ref{eq:USLIC}. 

A good approximation for the double-quantum excitation amplitude, for the scheme in Figure~\ref{fig:SLIC-DQ-sequences}(a), may be calculated by considering the effect of $U(T)$ on longitudinal magnetization:
\begin{align}
\rho_f(T) = U(T) I_z U(T)^{\dagger}
\end{align}
This leads to the following DQ excitation amplitude, denoted $a_{\rm{DQ}}^{\rm{SLIC(a)}}(T)$:
\begin{align}
\label{eq:aDQSLICwithepsilon}
a&_{\rm{DQ}}^{\rm{SLIC(a)}}(T) = \bra{2}\rho_f(T)\ket{4} 
\simeq 
\nonumber  \\ 
\tfrac12{i}
\bigg\{&\sin\left(\omega_{J}\left(1+\epsilon_{\rm{rf}}\right) T\right) 
\nonumber \\-  &\sin\left(\tfrac{1}{2} \omega_{J}\left(2+\epsilon_{\rm{rf}}\right)T \right)\cos\left(\tfrac{1}{2}\Omega_{12} T\right) 
\nonumber \\
- &\cos\left(\tfrac{1}{2} \omega_{J}\left(2+\epsilon_{\rm{rf}}\right) T\right)\sin\left(\tfrac{1}{2} \Omega_{12} T\right)\sin\left(\gamma_{\epsilon}\right) \bigg\}
\end{align} 

The equation for $a_{\rm{DQ}}^{\rm{SLIC}(a)}(T)$ is in good agreement with the numerical simulations shown in Figure~\ref{fig:SLIC-RFerrors-plot}, but subtly fails to reproduce some features such as the asymmetry about $\epsilon_{\rm{rf}} = 0$, which generally requires a consideration of counter-rotating Hamiltonian terms \cite{shirley_solution_1965, zhang_unveiling_2020}.

\section{Double-quantum excitation by cSLIC in the presence of rf amplitude errors}
\label{appx:cSLICtheory}

Here we evaluate the performance of the compensated SLIC sequence as a function of a fractional error $\epsilon_\mathrm{rf}$ in the rf field amplitude, causing a mismatch of the SLIC match condition (Equation~\ref{eq:SLICcondition}). Exact resonance is assumed ($\omega_\Sigma=0$).  The repeating element of the cSLIC pulse sequence, as given in Equation~\ref{eq:cSLIC}, is a time-symmetric sequence of three pulses, with durations $\{\tau_a,\tau_b,\tau_c\}$, phases $\{0,\pi,0\}$, and rf amplitudes denoted here by the nutation frequencies $\{\omega_a,\omega_b,\omega_c\}$. Taking into account the rf amplitude error, the nutation frequencies are given by
\begin{align}
 \omega_a    &= \omega_{J} (1+\epsilon_{\mathrm{rf}})
\nonumber \\
 \omega_b    &= \omega_{\mathrm{strong}} (1+\epsilon_{\mathrm{rf}})
 \nonumber \\
 \omega_c &= \omega_a
\end{align}
This assumes that the rf amplitude error $\epsilon_\mathrm{rf}$ affects the amplitude of all three pulses in equal proportion, which is the case if the amplifiers are linear and all pulses are generated by the same rf coil. 

The durations are given by 
\begin{align}
 \tau_a    &= \alpha\pi/\omega_J 
\nonumber \\
  \tau_b    &= \alpha2\pi/\omega_{\rm{strong}} =(1-\alpha)2\pi/\omega_{J} 
\nonumber \\
    \tau_c &= \tau_a
\end{align}
where $\alpha$ is defined in Equation~\ref{eq:alpha factor}. The 
nominal pulse flip angles are $\{\alpha\pi,2\alpha\pi,\alpha\pi\}$. The complete three-pulse sequence has duration
\begin{equation}
    \tau_a + \tau_b + \tau_c = \tau_J
\end{equation}

Now consider an interaction frame defined by the time-dependent operator $W(t)$, which takes into account the rf amplitude error $\epsilon_\mathrm{rf}$. For the cSLIC sequence, this may be written as follows: 
\begin{equation}
    W(t)=R_{y}(-\pi/2)U_J(t)R_z(-\varphi(t))
\end{equation}
where the angle $\varphi(t)$ is a piecewise function taking into account the angle accumulated by the three commuting rotations, with the central rotation in opposite sense to the others:
\begin{multline}
\varphi(t)=
\left\{
\begin{array}{ccc}
   \omega_a (t-t_a^0)
        &;\quad & 
    t_a^0 \leq t \leq t_b^0 
        \\[2pt]
   \omega_a \tau_a    
   - \omega_b (t-t_b^0)
        &;\quad & 
    t_b^0\leq t \leq t_c^0 
        \\[2pt]
   \omega_a\tau_a - \omega_b\tau_b + \omega_a(t-t_c^0)
        &;\quad & 
    t_c^0 \leq t \leq  \tau_J 
        \\[2pt]
\end{array}
\right.
\end{multline}
The initial time points of the three intervals are given by
\begin{align}
t_a^0 &=0
\nonumber\\
t_b^0 &=\tau_a
\nonumber\\
t_c^0 &=\tau_a + \tau_b
\end{align}

The accumulated rotation angle is identically zero at the end of the three-pulse sequence:
\begin{equation}
    \varphi(\tau_J)=0
\end{equation}
Hence, the interaction frame is cyclic even for finite $\epsilon_{\rm{rf}}$, with a period $\tau_J$:
\begin{equation}
\varphi(t)= \varphi(t+\tau_J) 
\end{equation}
This important property allows the average Hamiltonian of the spin interactions, expressed in the interaction frame, to be used for the analysis of an arbitrary number of cSLIC repetitions. 

The Hamiltonian for the chemical shift difference may be expressed in the interaction frame as follows:
\begin{equation}
\tilde{H}_{\Delta}(t) = 
W(t)^\dagger H_{\Delta} W(t) 
\end{equation}

In the limit of a negligibly short compensating pulse (the condition $\tau_b/\tau_a\rightarrow 0$, which is synonymous with $\alpha \rightarrow 1$), the average Hamiltonian $\overline{H}^{(1)}_{\Delta}$ of the cSLIC sequence, in the presence of rf errors, can be written as:
\begin{align}
\overline{H}^{(1)}_{\Delta} &= \lim_{\alpha \rightarrow 1} \int_0^{1/J} \tilde{H}_{\Delta}(t) \dd{t} \nonumber\\
&= 
-\kappa\omega_{\Delta }
\big[
\mathrm{sinc}\left(f_{+}\right) I_x^{12} - \mathrm{sinc}\left(f_{-}\right) I_x^{14}
\big]
\end{align} 
where $\kappa=2^{-1/2}$, the sinc function takes the usual definition ($\mathrm{sinc}(x)=x^{-1}\sin x$), and the arguments of the $\mathrm{sinc}$ functions are:
\begin{align}
f_{+} &= \pi \epsilon_{\rm{rf}} \nonumber \\
f_{-} &= \pi \left(2+\epsilon_{\rm{rf}}\right)
\end{align}
These represent a pair of counter-rotating frequency components, centred at the resonances of the two nominal SLIC matching conditions: $\omega_{\rm{nut}}^{\rm{SLIC}} = +\omega_J$ ($\epsilon_{\rm{rf}} =0$), and $\omega_{\rm{nut}}^{\rm{SLIC}} = -\omega_J$ ($\epsilon_{\rm{rf}} =-2$).

This suggests that $\overline{H}^{(1)}_{\Delta}$ can be associated with a modified scaling factor $\kappa'$ and detuning angle $\theta_{\epsilon}$, defined as follows:
\begin{align} 
\theta_{\epsilon}
&= 2\,\mathrm{arctan}\left(f_{+}/f_{-}\right) \nonumber \\ \kappa' &=  
\kappa \ \mathrm{sinc}\left(f_{+} \right)\sec\left(\theta_{\epsilon}\right) 
\end{align}
These expressions are equivalent to Equation \ref{eq:thetaepsilon+kappap}. If there is no rf amplitude deviation ($\epsilon_\mathrm{rf}=0$), the detuning angle vanishes ($\theta_\epsilon=0$) and the scaling factor becomes $\kappa'=\kappa=2^{-1/2}$.

With these definitions, $\overline{H}^{(1)}_{\Delta}$ can be recast in terms of a $I_{x}^{12}$ single-transition operator that has been rotated about the y-axis of the $\{\ket{2},\ket{4}\}$ subspace by an angle $\theta_{\epsilon}$:
\begin{align}
\overline{H}^{(1)}_{\Delta}  
&= -\kappa'\omega_{\Delta} 
\left[ \cos\left(\tfrac{1}{2}\theta_{\epsilon}\right) I_{x}^{12}-\sin\left(\tfrac{1}{2}\theta_{\epsilon}\right)I_{x}^{14} \right] \nonumber \\
&= -\kappa'\omega_{\Delta}
R_y^{24}\!\left(\theta_{\epsilon}\right)^{\dagger}
I_{x}^{12}
R_y^{24}\!\left(\theta_{\epsilon}\right)
\end{align}

This formulation of $\overline{H}^{(1)}_{\Delta}$ permits the associated effective propagator, $\overline{U}_{\Delta}(t)$, to be expressed as a product of rotations in the $\{\ket{1},\ket{2}\}$ and $\{\ket{2},\ket{4}\}$ subspaces:
\begin{align}
\overline{U}_{\Delta}(t) &= \exp\{-i \overline{H}_{\Delta}^{(1)}t\} \nonumber \\ &=R_y^{24}\!\left(\theta_{\epsilon}\right)^{\dagger} R_x^{12}\!\left(-\kappa'\omega_{\Delta} t\right)R_y^{24}\!\left(\theta_{\epsilon}\right)
\end{align}

The effective propagator $\overline{U}_{\Delta}(t)$, when applied for a total duration $T$ and followed by a $(\pi/2)_x$ pulse, transforms in-phase transverse magnetization $I_x$ into a final density operator designated $\rho_f(T)$, which contains double-quantum coherence:
\begin{equation}
\rho_f(T) = R_x(\pi/2)\overline{U}_{\Delta}(T) I_x\overline{U}_{\Delta}^{\dagger}(T)R_x^{\dagger}(\pi/2)
\end{equation}

The double-quantum excitation amplitude may be extracted from $\rho_{f}(T)$ as follows:
\begin{align}
 a_{\mathrm{DQ}}^\mathrm{cSLIC} (T)  &=  
 \bra{2} \rho_f(T)\ket{4}
 \nonumber \\ &=  
 -i \cos\left(
 \theta_{\epsilon}
 \right)
 \sin^2\!\left(
 \tfrac{1}{4}
 \kappa'\omega_{\Delta} T\right) 
\end{align}

The enhanced performance of the cSLIC sequence over the SLIC sequence can be explained by comparing the series expansions of their respective detuning angles $\theta_{\epsilon}$ and $\gamma_{\epsilon}$:

\begin{align}
\theta_{\epsilon} &\simeq \epsilon_{\rm{rf}}+\mathcal{O}(\epsilon_{\rm{rf}}^2) \nonumber \\
\gamma_{\epsilon}&\simeq \sqrt{2}\epsilon_{\rm{rf}}J/\Delta+\mathcal{O}(\epsilon_{\rm{rf}}^2) \nonumber
\end{align}

This shows that the double-quantum excitation induced by the cSLIC sequence is independent of the undesirable factor $J/\Delta$.

The performance of the cSLIC sequence, with respect to the finite duration of the compensating pulse (i.e. when $\alpha \neq 1$), is expected to be unaffected so long as $\alpha \ge 0.95$, corresponding to the practical constraint that the nutation frequency of the compensating pulse should be at least roughly $\sim 20$ times the J-coupling. This behavior is shown in simulations in the Supporting Information.

\section*{Supplementary material}
The supplementary material contains: 
Details of the synthesis of \textbf{I};
control double-quantum filtering experiments; the theory of PulsePol/Symmetry-Based DQ excitation using incomplete $RN_n^\nu$ cycles; 
finite pulse effects on the cSLIC sequence; estimation of the \textsuperscript{19}F $T_1$ relaxation time of \textbf{I} in solution.

\section*{Acknowledgments}
We acknowledge funding from the European Research Council (Grant No. 786707-FunMagResBeacons) and EPSRC-UK (Grant Numbers EP/P030491/1, EP/V047663/1, EP/V055593/1, and EP/V047663/1). We thank Piotr Garbacz for useful discussions on DQ NMR in strongly-coupled $^{19}$F spin systems. We thank Matthew S. Rosen and Stephen DeVience for fruitful discussions on the SLIC pulse sequence, and Nino Wili for valuable insight in pulse sequence design. We are grateful to Sharon Ashbrook for discussions, and Marek Plata for instrumental support.

\section*{Author declarations}
\subsection*{Conflicts of interest}
The authors have no conflicts to disclose.

\subsection*{Author Contributions}

\textbf{Urvashi D. Heramun}: Investigation (equal); Conceptualization (supporting); Data curation (equal); Visualization (equal); Validation (equal); Formal analysis (equal); Writing – original draft (equal); Writing – review \& editing (equal).

\textbf{Mohamed Sabba}: Investigation (equal); Conceptualization (lead); Data curation (equal); Visualization (equal); Methodology (equal); Validation (equal);  Formal analysis (equal); Writing – original draft (equal); Writing – review \& editing (equal).

\textbf{Dolnapa Yamano}: 
Conceptualization (supporting); Resources (lead); Methodology (equal); Investigation (equal); Writing – original draft (supporting).

\textbf{Christian Bengs}: Conceptualization (supporting); Methodology (supporting); Writing – review \& editing (supporting).

\textbf{Bonifac Legrady}: Conceptualization (supporting); Methodology (supporting).

\textbf{Giuseppe Pileio}: 
Conceptualization (supporting).

\textbf{Sam Thompson}: 
Conceptualization (equal); Resources (equal); Funding acquisition (equal); Writing – review \& editing (supporting).

\textbf{Malcolm H. Levitt}: Conceptualization (equal); Data curation (equal); Formal analysis (equal); Funding acquisition (equal); Validation (equal); Project administration (lead); Software (equal); Supervision (lead); Writing – review \& editing (equal); Writing – original draft (equal).

\section*{Data availability}

The data that support the findings of this study are available from the corresponding author upon reasonable request. The pulse sequences and SpinDynamica notebook are publicly available with the assigned DOI: https://doi.org/10.5258/SOTON/D3634.

\bibliography{References/UDH-spinorDQ}
\end{document}


\title{Spinor Double-Quantum Excitation in the Solution NMR of Near-Equivalent Spin-1/2 Pairs}

\author{Urvashi D. Heramun}
\affiliation{School of Chemistry, University of Southampton, SO17 1BJ, UK}

\author{Mohamed Sabba}
\affiliation{School of Chemistry, University of Southampton, SO17 1BJ, UK}

\author{Dolnapa Yamano}
\affiliation{School of Chemistry, University of Southampton, SO17 1BJ, UK}

\author{Christian Bengs}
\affiliation{School of Chemistry, University of Southampton, SO17 1BJ, UK}

\author{Bonifac Legrady}
\affiliation{School of Chemistry, University of Southampton, SO17 1BJ, UK}

\author{Giuseppe Pileio}
\affiliation{School of Chemistry, University of Southampton, SO17 1BJ, UK}

\author{Sam Thompson}
\affiliation{School of Chemistry, University of Southampton, SO17 1BJ, UK}

\author{Malcolm H. Levitt}
\email[]{mhl@soton.ac.uk}
\affiliation{School of Chemistry, University of Southampton, SO17 1BJ, UK}

\date{\today}

\maketitle

\tableofcontents

\renewcommand{\hl}[1]{{#1}}
\section{Supplementary Material}
\subsection{Synthesis of \textbf{I}}
\subsubsection{Solvents and Reagents}
Reactions were carried out under an argon atmosphere in oven-dried glassware at room temperature unless otherwise stated. Standard inert atmosphere techniques were used in handling all air- and moisture-sensitive reagents. All reagents were purchased from Sigma-Aldrich, ThermoFisher Scientific, or Fluorochem and used without further purification. Anhydrous and deuterated solvents were purchased from Merck or Fluorochem and were used as supplied. Aqueous solutions are saturated unless specified otherwise.

\subsubsection{Chromatography}
Flash column chromatography was carried out using Merck $60$ silica gel. Thin-layer chromatography was carried out using Merck Kieselgel $60$ F254 ($230$-$400$ mesh) fluorescent treated silica, visualized under UV light ($254$ nm) or by staining with aqueous potassium permanganate solution, ninhydrin or ceric ammonium molybdate solutions.

\subsubsection{Spectroscopy}
\textsuperscript{1}H, \textsuperscript{13}C, and \textsuperscript{19}F spectra were recorded using a Bruker spectrometer ($500$ MHz for \textsuperscript{1}H, $126$ MHz for \textsuperscript{13}C and $471$ MHz for \textsuperscript{19}F) running TopSpin\texttrademark software and are quoted in ppm for measurement against residual solvent peaks (\textsuperscript{1}H, \textsuperscript{13}C) or externally referenced against CFCl\textsubscript{3} in the appropriate solvent (\textsuperscript{19}F). Chemical shifts ($\delta$) are given in parts per million (ppm) and coupling constants ($J$) are given in Hertz (Hz). The \textsuperscript{1}H NMR spectra are reported as follows: $\delta$ (number of protons, multiplicity, coupling constant). Multiplicity is abbreviated as follows: s = singlet, d = doublet, t = triplet, q = quartet, quint. = quintet, m = multiplet, br = broad. Systematic compound names are those generated by ChemDraw\texttrademark ~(CambridgeSoft) following IUPAC nomenclature. The numbering scheme used for NMR assignment is arbitrary and does not follow any convention. Numbering of compounds is illustrated on the spectra themselves, \textit{vide infra}. The \textsuperscript{13}C and \textsuperscript{19}F NMR spectra are reported in $\delta$ / ppm. \textsuperscript{13}C and \textsuperscript{19}F NMR spectra are proton decoupled unless specified otherwise. Where necessary, two-dimensional (COSY, HSQC, HMBC) NMR experiments were used to assist the assignment of signals in the \textsuperscript{1}H and \textsuperscript{19}F NMR spectra.

High-resolution ESI mass spectra were undertaken at the University of Southampton using a MaXis (Bruker Daltonics, Bremen, Germany) time of flight (TOF) mass spectrometer or a solariX (Bruker Daltonics, Bremen, Germany) mass spectrometer equipped with a $4.7$ T magnet and FT-ICR cell. 

Infra-red (IR) spectra were recorded using a ThermoFisher Scientific Nicolet iS5 spectrometer from a thin film deposited onto a diamond ATR module. Sixteen scans were acquired at a resolution of $4$ cm\textsuperscript{-1}. Only selected maximum absorbances ($\nu_{\text{max}}$) of the most intense peaks are reported (cm\textsuperscript{-1}).

Melting points were recorded using a Stuart (Bibby Scientific Limited) SMP20 machine and reported uncorrected in degrees Celsius ($\degree$C).

Optical rotations were recorded on an Optical Activity POLAAR $2001$ at $589$ nm.

\begin{figure*}[tbh]
\centering
\includegraphics[width=0.75\linewidth]{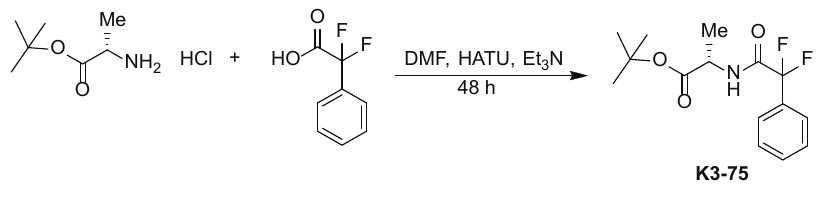}
\caption{
Compound synthesis. 
}
\label{fig: 1}
\end{figure*}

\begin{figure*}[tbh]
\centering
\includegraphics[width=0.5\linewidth]{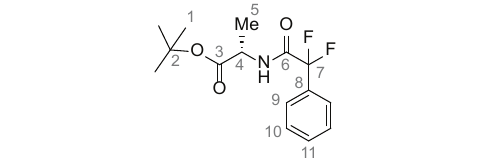}
\caption{
Structure of the fluorinated amide: tert-butyl (2,2-difluoro-2-phenylacetyl)-L-alaninate, also called \textbf{K3-75}.
}
\label{fig: 2}
\end{figure*}

To a solution of 2,2-difluoro-2-phenylacetic acid ($98$ mg, $0.57$ mmol) in DMF ($2.0$ mL), Et\textsubscript{3}N ($0.20$ mL, $1.43$ mmol) was added, and the mixture was stirred for $5$ minutes. HATU ($261$ mg, $0.69$ mmol) was added, and the reaction mixture stirred for $10$ minutes. L-Alanine \textit{tert}-butyl ester hydrochloride ($104$ mg, $0.57$ mmol) was added. The reaction was stirred for 18 hours. Saturated NaHCO\textsubscript{3} ($10 mL$) was added followed by extraction with EtOAc ($2 \times 10$ mL). The organic layer was washed with brine ($3 \times 10$ mL), dried over magnesium sulfate, filtered, and concentrated in vacuo. The residue was purified by flash column chromatography (ethyl acetate/petroleum ether, 1/4) to give the title compound K3-75 ($123$ mg, $0.41$ mmol, $72\%$ yield) as a white solid.

R\textsubscript{f} $0.40$ (ethyl acetate/petroleum ether, 1/4).

[$\alpha$]\textsubscript{D}\textsuperscript{25} – 4.0 (\textit{c} $0.67$, CH\textsubscript{2}Cl\textsubscript{2}).

MP $48-50$ $\degree$C.

$\delta_{\text{H}}$ ($500$ MHz, CDCl\textsubscript{3}) $7.63-7.61$ ($2$H, m, H$9$), $7.49-7.43$ ($3$H, m, H$10$, H$11$), $7.1$ ($1$H, s, NH), $4.45$ ($1$H, p, $J$ $7.1$, H$4$), $1.46$ ($9$H, s, H$1$), $1.43$ ($3$H, d, $J$ $7.1$, H$5$).

$\delta_{\text{C}}$ ($126$ MHz, CDCl\textsubscript{3}) $171.3$ (C$3$), $163.5$ (t, $J$ 31.5, C$6$), $133.1$ (t, $J$ 25.4, C$8$), $131.0$ (t, $J$ $1.7$, C$10$), $128.7$ (C$11$), $125.7$ (t, $J$ 6.2, C$9$), $114.8$ (t, $J$ 252.8, C$7$), $82.9$ (C$2$), $49.0$ (C$4$), $28.0$ (C$1$), $18.4$ (C$5$).

$\delta_{\text{F}}$ ($471$ MHz, CDCl\textsubscript{3}) $-102.9$.

$\delta_{\text{F}}$ \{\textsuperscript{1}H\} ($471$ MHz, CDCl\textsubscript{3}) $-102.9$.

$\nu_\text{max}$ $3378$, $3001$, $2981$, $2945$, $2922$, $1722$, $1686$, $1532$, $1452$, $1367$, $1318$, $1261$, $1247$, $1153$, $1091$, $1063$, $1003$, $938$.

HRMS (ESI) found [M$+$Na]\textsuperscript{+} $322.1225$, C\textsubscript{15}H\textsubscript{19}F\textsubscript{2}NNaO\textsubscript{3} requires $322.1225$.

\begin{figure*}[tbh]
\centering
\includegraphics[width=\linewidth]{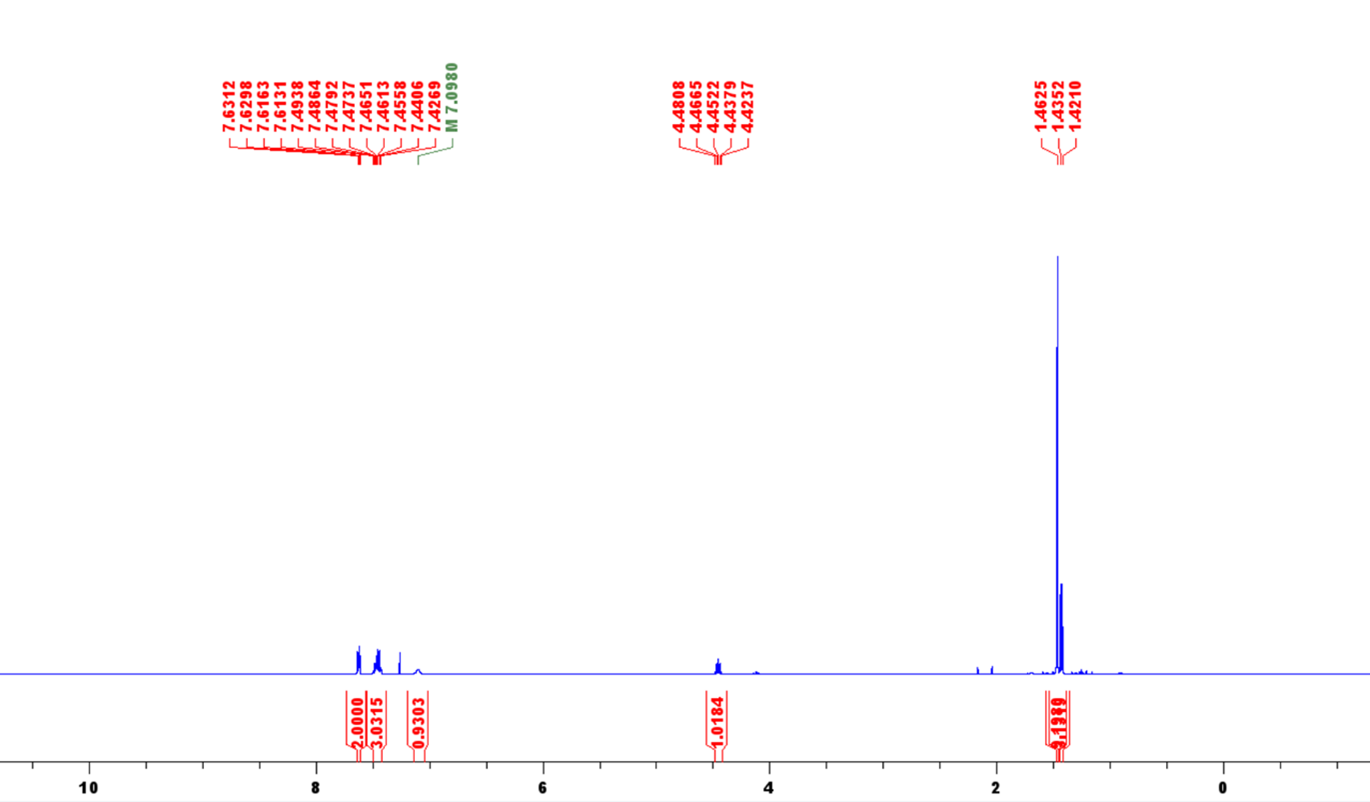}
\caption{
\textsuperscript{1}H, $500$ MHz, CDCl\textsubscript{3} spectrum. 
}
\label{fig: 3}
\end{figure*}

\begin{figure*}[tbh]
\centering
\includegraphics[width=\linewidth]{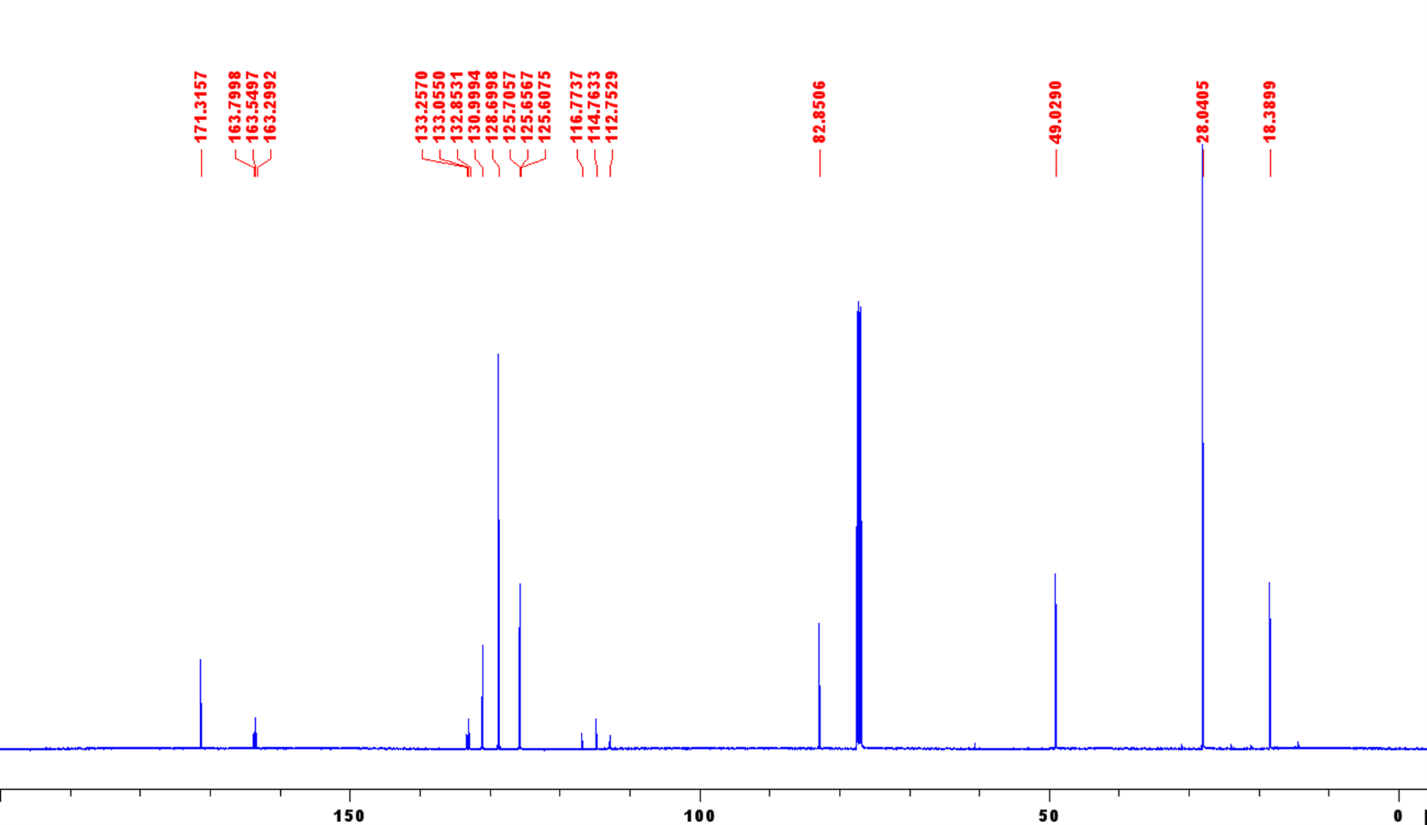}
\caption{
\textsuperscript{13}C, $126$ MHz, CDCl\textsubscript{3} spectrum. 
}
\label{fig: 4}
\end{figure*}

\begin{figure*}[tbh]
\centering
\includegraphics[width=\linewidth]{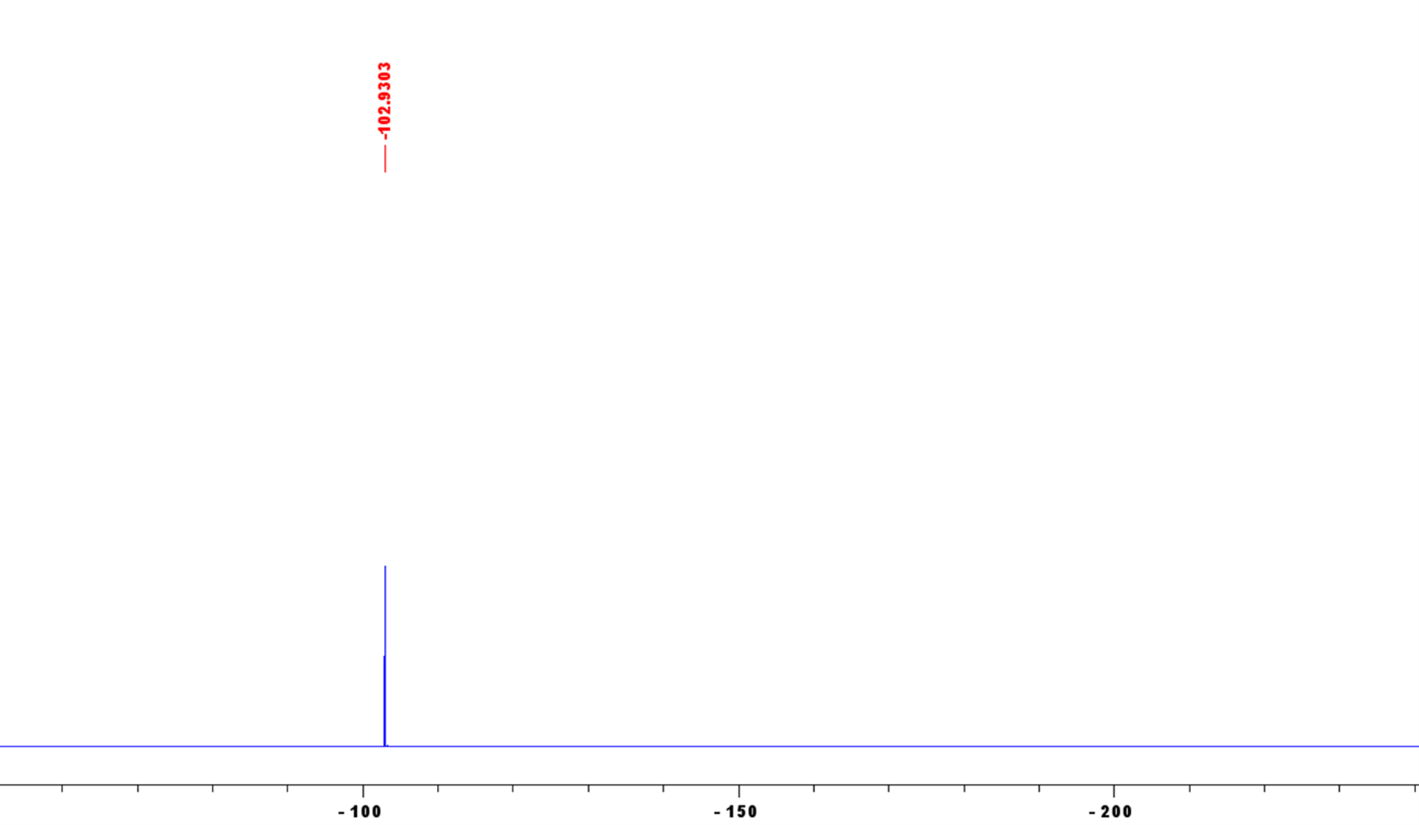}
\caption{
\textsuperscript{19}F, $471$ MHz, CDCl\textsubscript{3} spectrum. 
}
\label{fig: 5}
\end{figure*}

\begin{figure*}[tbh]
\centering
\includegraphics[width=\linewidth]{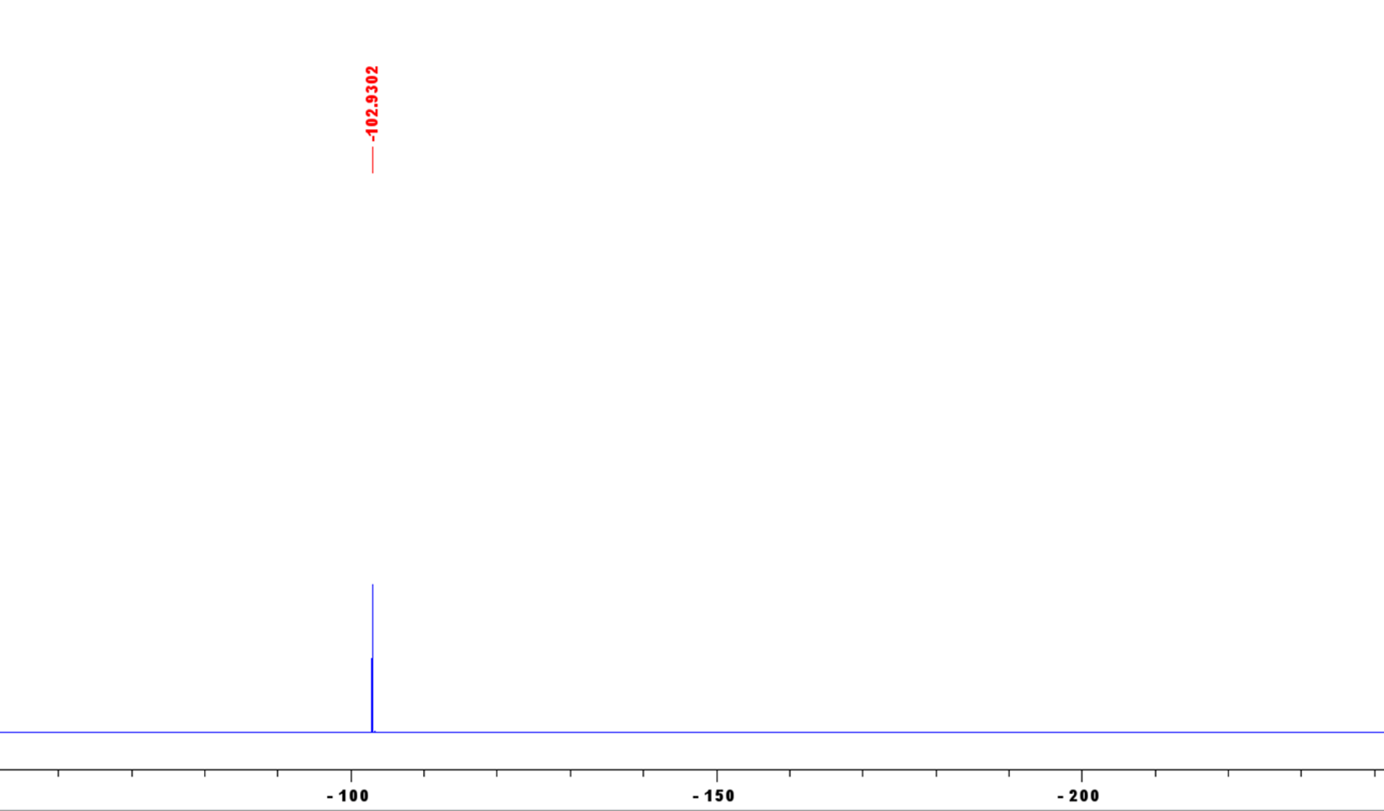}
\caption{
\textsuperscript{19}F \{\textsuperscript{1}H\}, $471$ MHz, CDCl\textsubscript{3} spectrum. 
}
\label{fig: 6}
\end{figure*}

\clearpage
\subsection{Control DQF Experiments}
\begin{figure}[tbp]
\centering
\includegraphics[width=0.35\textwidth]{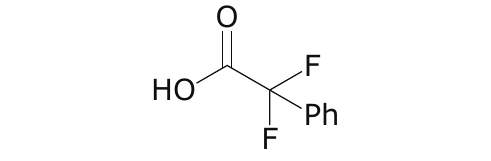}
\caption{Molecular structure of 2,2-difluoro-2-phenylacetic acid. The two \F{19} nuclei are magnetically equivalent. This system was used as a control for the operation of the double-quantum filter. 
}
\label{fig:control-molecule}
\end{figure}

\begin{figure}[btp]
\centering
\includegraphics[width=0.5\linewidth]{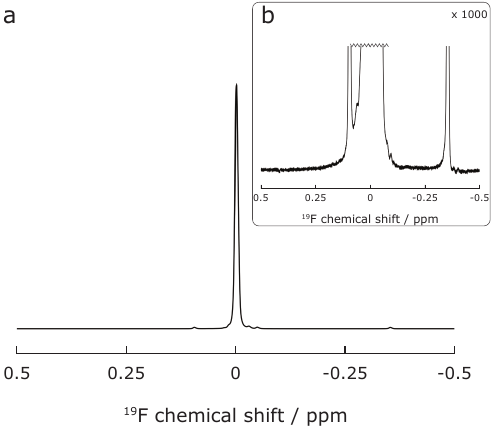}
\caption{
(\textbf{a}) A $90\degree$ pulse-acquire \textsuperscript{19}F NMR spectrum
of a solution of the fluorinated carboxylic acid molecule shown in Figure~\ref{fig:control-molecule} in CDCl\textsubscript{3},
acquired at 14.1 T, 298 K, and averaged over 5120 transients. The frequency scale is centred around $-105.148$ ppm. (\textbf{b}) A zoomed in view of the spectrum shown in part a. The vertical scale is given.
}
\label{fig:ControlSpectra}
\end{figure}

To verify that the double-quantum filtered spectra shown in the main text arise due to the chiral-induced symmetry-breaking mechanism, \enquote{control} experiments were performed on a solution of a fluorinated carboxylic acid without the chiral substituent. The molecular structure is shown in Figure~\ref{fig:control-molecule}. 
Since the two \F{19} nuclei are magnetically equivalent, the chemical shift difference $\Delta\delta$ is zero, and the \F{19}-\F{19} $J$-coupling is unobservable.
In a $90\degree$ pulse-acquire spectrum averaged over $5120$ transients, the AB pattern characteristic of a strongly-coupled system is absent (Figure~\ref{fig:ControlSpectra}). Attempts at pulse sequence parameter optimization were nonetheless carried out (see Table~\ref{table: ControlPulseSequenceParameters}).

\begin{figure*}[htbp]
\centering
\includegraphics[width=\linewidth]{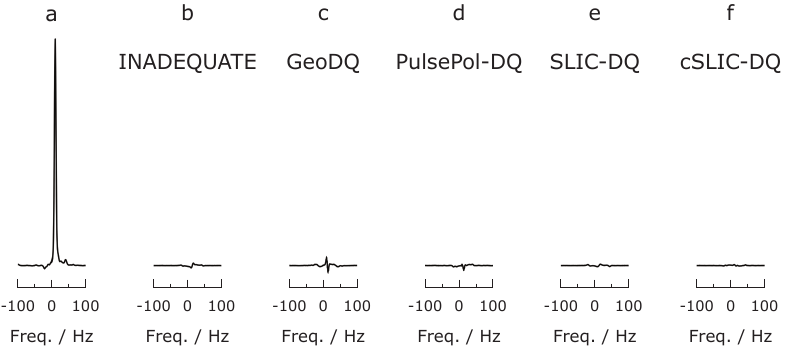}
\caption{
\F{19} spectra of a solution of the fluorinated carboxylic acid molecule shown in Figure~\ref{fig:control-molecule} in CDCl\textsubscript{3}, acquired at $14.1$ T, $298$ K, and averaged over 4 transients. $\mathrm{^1H}$ decoupling was not used. The frequency axis is centred around $-105.148$ ppm. All spectra have the same vertical scale. All pulse sequence parameters are given in table~\ref{table: ControlPulseSequenceParameters}.
(\textbf{a}) A $90\degree$ pulse-acquire spectrum.
(\textbf{b}) Double-quantum filtered spectrum obtained using the refocused INADEQUATE pulse sequence. 
(\textbf{c}) Double-quantum filtered spectrum obtained using the GeoDQ pulse sequence. 
(\textbf{d}) Double-quantum filtered spectrum obtained using the PulsePol-DQ pulse sequence.
(\textbf{e}) Double-quantum filtered spectrum obtained using the SLIC-DQ pulse sequence. 
(\textbf{f}) Double-quantum filtered spectrum obtained using the cSLIC-DQ pulse sequence. 
}
\label{fig:control-DQF-spectra}
\end{figure*}

\begin{table}[htbp]
\begin{center}
\caption{Pulse sequence parameters used in the control double-quantum filtering experiments.}
\renewcommand{\arraystretch}{1.5}
\begin{tabular}{c c c}  
\hline
 \textbf{Pulse Sequence} & \textbf{Parameter} & \textbf{Value} \\ \hline
Hard pulses & $\omega_\mathrm{nut}/(2\pi)$ / kHz & $22.8 \pm 0.1$ 
\\ \hline
INADEQUATE & $\tau_1$ / ms & $500.00$ 
\\ \hline
\multirow{3}{*}{GeoDQ} & $\tau_{1}^\mathrm{Geo}$ / $\mu$s & $976.0$ \\ 
& $\tau_{2}^\mathrm{Geo}$ / $\mu$s & $1952$ \\
& $n$ & $26$ 
\\ \hline 
\multirow{2}{*}{PulsePol-DQ} 
& $\tau_{2}$ / $\mu$s 
& $2400$ \\
& $m$ & $9$
\\ \hline 
SLIC-DQ & $\omega_\mathrm{nut}^\mathrm{SLIC}/(2\pi)$ / Hz & $226  \pm 3$
\\ \hline
\multirow{4}{*}{\shortstack{cSLIC-DQ}}
& {$\omega_\mathrm{nut}^\mathrm{SLIC}/(2\pi)$} / Hz & $256 \pm 3$ \\
& $n_J$ & $21$ \\ 
& $\tau_J$ / $\mu$s & $3940$
\\
& $\alpha$ & $-$
\\
\hhline{= = =}
\end{tabular}
\label{table: ControlPulseSequenceParameters}
\end{center}
\end{table}

\clearpage

\subsection{Double-Quantum Excitation With Incomplete Cycles}
\label{appx:IncompleteRNnnu}

If the PulsePol/Symmetry-Based variant of Spinor-DQ excitation is used, but without completing a whole number of $RN_n^\nu$ cycles, the propagators $U_\mathrm{rf}$ and $U_J$ in 
Equation 48 of the main text must be taken into account. The propagator $U_J$ commutes with the double-quantum precursor state $\rho_3$ in
Equation 33 of the main text,
and has no influence on the result, even for incomplete $RN_n^\nu$ cycles. However, the rf propagator $U_\mathrm{rf}(t)$ has a strong effect on the double-quantum excitation, unless the excitation sequence is terminated at a carefully-chosen time point, or additional precautions are taken.

The general case is worth careful consideration, since the requirement to complete entire $RN_n^\nu$ cycles strongly restricts the available values of the double-quantum excitation time. In general, a pulse sequence with the symmetry $RN_n^\nu$ consists of $N$ concatenated $R$-elements, with phases alternating between the values $\pm \pi\nu/N$. Suppose that $n_R$ $R$-elements are completed, corresponding to a total excitation time
$T=(n_R n)/(N J)$.
If $n_R$ is even, the resulting rf propagator is given by: 
\begin{equation}
U_\mathrm{rf}(n_R)
=
[
R_{-\phi}(\pi)
R_{+\phi}(\pi)
]^{n_R/2},
\end{equation}
where:
\begin{equation}
R_\phi(\pi) = R_z(\phi)R_x(\pi)R_z(-\phi),
\end{equation}
and:
\begin{equation}
\phi=\pi\nu/N.
\end{equation}
The overall rf propagator evaluates to:
\begin{align}
\label{eq:Urf(nR)}
U_\mathrm{rf}(n_R)
&=
[
R_z(-\phi)R_x(\pi)R_z(+\phi)
\nonumber\\
&\qquad \times 
R_z(+\phi)R_x(\pi)
R_z(-\phi)
]^{n_R/2}
\nonumber\\
&=
[
R_z(-2\phi)R_x(\pi)
R_x(\pi)R_z(-2\phi)
]^{n_R/2}
\nonumber\\
&=
R_z(-2 n_R \phi)
R_x(2\pi n_R )
\nonumber\\
&=
R_z(-2\pi n_R\nu/N)
R_x(2\pi n_R ).
\end{align}
The final operator in Equation~\ref{eq:Urf(nR)} denotes a rotation through a multiple of $2\pi$ and has no effect on the final state. If $n_R$ is a multiple of $N$ (indicating complete $RN_n^\nu$ cycles), the first term on the last line also indicates a rotation through a multiple of $2\pi$ and has no effect. This confirms that the propagator $U_\mathrm{rf}$ is harmless for the double-quantum excitation, if complete $RN_n^\nu$ cycles are performed.

Consider now the case in which an odd number of half-cycles is completed:
\begin{equation}
    n_R = (2k+1)N/2; 
\quad k \in \mathbb{Z}.
\end{equation}
In this case, the relevant operator evaluates to:
\begin{equation}
R_z(-2\pi n_R\nu/N)
=
R_z(-\pi (2k+1) \nu) ;
\quad k \in \mathbb{Z}.
\end{equation}
In this case, $U_\mathrm{rf}$ gives rise to a rotation about the $z$-axis through an odd multiple of $\pi$ in the case that $\nu$ is odd. This rotation changes the sign of the state $\rho_3$ in
Equation 33 of the main text,
and hence changes the sign of the excited double-quantum coherence. This is usually not harmful, since the sign change is reversed when the double-quantum coherence is converted back into observable magnetization by a similar pulse sequence. Hence, it is acceptable to truncate the DQ excitation sequence half-way through the last $RN_n^\nu$ sequence, if the sign of the excited double-quantum coherence is unimportant. In the case of the $R4_3^1$ sequence in 
Equation 43 of the main text,
any integer number of phase-shifted PulsePol sequences may therefore be used for DQ excitation.  

The sign alternation of the double-quantum amplitude with respect to the number of half-cycles is apparent in the simulation shown in 
Figure 8(a) of the main text.

Further subdivision of the $RN_n^\nu$ cycles is not compatible with double-quantum excitation, unless additional pulses and/or phase shifts are inserted, to correct for the left-over $U_\mathrm{rf}$ transformation.

\clearpage
\subsection{Finite Pulse Effects on the cSLIC Sequence}

Recall the first-order average Hamiltonian of the cSLIC sequence, as given in Equation C.13 of the main text:

\begin{equation}
\overline{H}^{(1)}_{\Delta}  
= -\kappa'\omega_{\Delta} 
\left[ \cos\left(\tfrac{1}{2}\theta_{\epsilon}\right) I_{x}^{12}-\sin\left(\tfrac{1}{2}\theta_{\epsilon}\right)I_{x}^{14} \right]
\end{equation}

Under this average Hamiltonian, Equation C.16 of the main text provides a definition of $a_{\rm{DQ}}^{\rm{cSLIC}}(T)$, the double-quantum excitation amplitude for cSLIC. For a nominal duration $T = \sqrt{2}/\Delta$, the squared modulus of Equation C.16 is:

\begin{equation}
\label{equation:cSLICamplitude}
\left\vert a_{\rm{DQ}}^{\rm{cSLIC}}\right\vert^2 = \cos^2\left(\theta_{\epsilon}\right)\sin^4\left(\tfrac{\pi}{2} \kappa'/\kappa\right)
\end{equation}

Where $\kappa = 2^{-1/2}$. Equation C.12 of the main text provides expressions for the pair of parameters $\theta_{\epsilon}$ and $\kappa'$ under the assumption that, in cSLIC, the length of the compensating pulse is negligible with respect to the length of the SLIC pulse. Outside of this assumption, when considering finite pulse effects, these parameters acquire a dependence on $\alpha$ (given in Equation 101 of the main text) as follows:

\begin{equation}
\label{equation:thetaexact}
\theta_{\epsilon} = \arctan\left[\frac{(\pi +\alpha  f_{+})(\pi -f_{-}) \text{sinc}(\alpha  f_{-})}{(\pi -\alpha  f_{-})(\pi +f_{+}) \text{sinc}(\alpha  f_{+})}\right].
\end{equation}

\begin{equation}
\label{equation:kappaexact}
\kappa' = \alpha\left[\frac{1}{2}\left(\frac{(\pi +f_{+})^2 \text{sinc}(\alpha f_{+} )^2}{(\pi +\alpha f_{+} )^2}+\frac{(\pi -f_{-})^2 \text{sinc}(\alpha f_{-})^2}{(\pi
   -\alpha f_{-})^2}\right)\right]^{1/2}.
\end{equation}

Where $f_{\pm}$ are given in Equation C.11 of the main text. The reader may verify that these expressions are equivalent to Equation C.12 of the main text in the limit $\alpha \rightarrow 1$.

\begin{figure*}[htbp]
\centering
\includegraphics[width=0.75\linewidth]{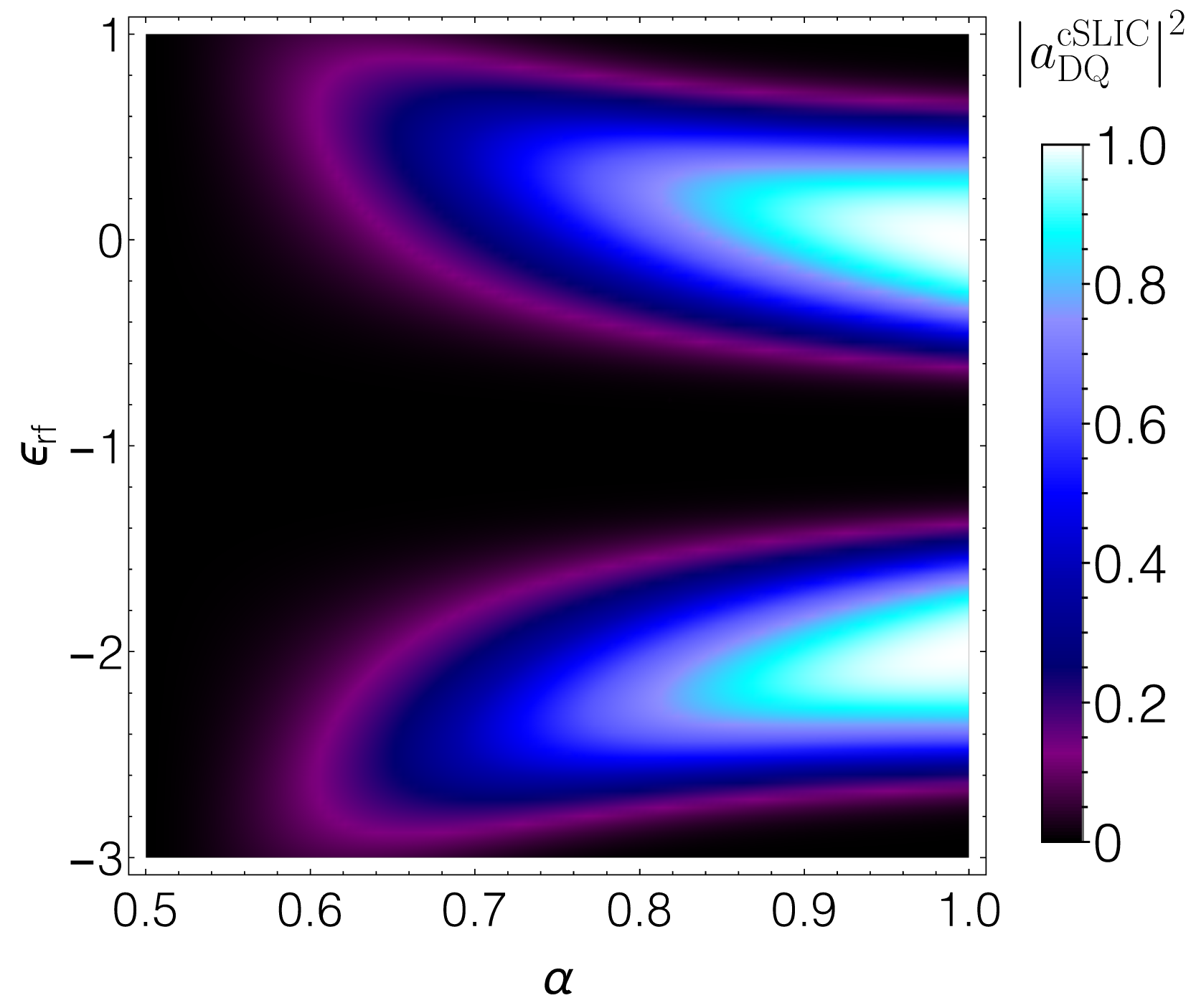}
\caption{Density plot of the double-quantum excitation efficiency of the cSLIC sequence to first order in average Hamiltonian theory (see Equation \ref{equation:cSLICamplitude}), assuming the generalized, $\alpha$-dependent values of $\kappa'$ and $\theta_{\epsilon}$ given in Equations~\ref{equation:thetaexact} and \ref{equation:kappaexact}. 
}
\label{fig:finitepulseeffects}
\end{figure*}

A density plot of Equation~\ref{equation:cSLICamplitude}, incorporating these expressions to predict the effect of $\alpha$ on double-quantum excitation is shown in Figure~\ref{fig:finitepulseeffects}. 
The plot, with a striking "butterfly wing" appearance, shows that so long as $\alpha \gtrapprox 0.95$, which roughly corresponds to a pulse power ratio of $\omega_{\rm{nut}}^{\rm{high}}/\omega_J \gtrapprox 20$, double-quantum excitation is expected to perform rather well. In practical terms, this may be understood as the minimum relative pulse strength required for the compensating pulse in cSLIC.

We remind the reader that $\epsilon_{\rm{rf}}$ is related to the nutation frequency of the nominal SLIC pulse, $\omega_{\rm{nut}}^{\rm{low}}$, as follows:
\begin{equation}
\epsilon_{\rm{rf}} = \left\{-2,-1,0\right\} \rightarrow \omega_{\rm{nut}}^{\rm{low}} =\left\{-\omega_{J},0,+\omega_J\right\}
\end{equation}
The two maximal regions in Figure~\ref{fig:finitepulseeffects} correspond to the SLIC conditions $\omega_{\rm{nut}}^{\rm{low}} =\pm\omega_J$.

\clearpage
\subsection{Experimental $\tone$ Relaxation Curve}

\begin{figure*}[htbp]
\centering
\includegraphics[width=0.75\linewidth]{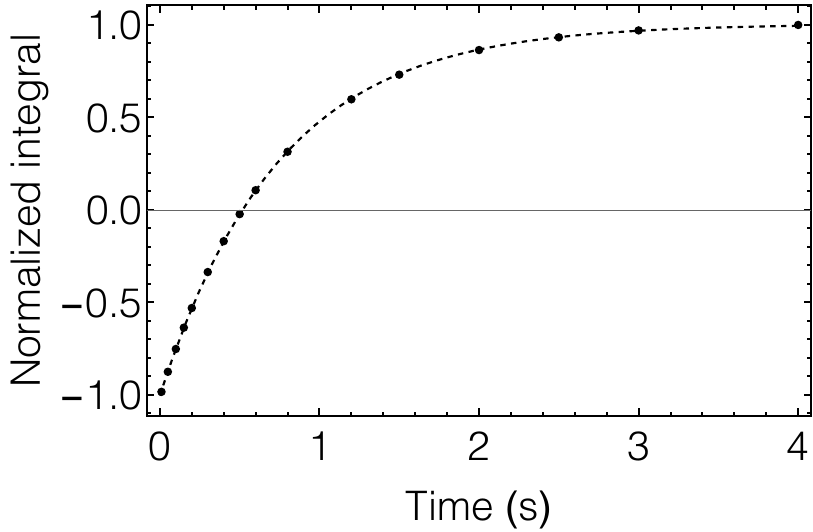}
\caption{Plot showing the relaxation of the spin system studied in the paper following an inversion recovery sequence.
Black points: experimental data. Dashed line: numerical fit to the function $f(t) = A - B \exp (-t/T_1)$. The numerical fit has the parameters $A = 1.005 \pm 0.002$, $B = 2.009 \pm 0.002$, and $T_1 = 746.8 \pm 0.3$ ms.}
\label{fig:t1curve}
\end{figure*}

A plot of the \F{19} relaxation of the spin system \textbf{I} investigated in the paper, obtained using a standard inversion recovery pulse sequence, is shown in Figure~\ref{fig:t1curve}. 

